\documentclass[11pt,a4paper]{article}
\pdfoutput=1
\usepackage{jheppub} 
\usepackage{amsmath}
\usepackage{amssymb}
\usepackage{feynman}

\newcommand{\gev}{{\rm GeV}}

\newcommand{\cL}{{\cal L}}
\newcommand{\cO}{{\cal O}}
\newcommand{\cM}{{\cal M}}
\newcommand{\cE} {{\cal E}}
\newcommand{\ztwo}{\mathbb{Z}_2}

\newcommand{\ratio}{{\frac{\gamma_w}{l_w}}}
\newcommand{\im}{{\rm Im}}

\newcommand{\beq}{\begin{eqnarray}}
\newcommand{\eeq}{\end{eqnarray}}

\subheader{\footnotesize
\vspace*{-3.7em}
\begin{flushright}
CERN-TH-2016-173
\end{flushright}
}

\title{Baryogenesis and Gravitational Waves from Runaway Bubble Collisions}

\author[a,b]{Andrey~Katz,}
\author[b]{and Antonio~Riotto}

\affiliation[a]{Theory Division, CERN, CH-1211 Geneva 23, Switzerland}
\affiliation[b]{
D\'epartement de Physique Th\'eorique and Center for Astroparticle Physics (CAP), \\ 
Universit\'e de Gen\`eve, 24 quai Ansermet, CH-1211 Gen\`eve 4, Switzerland}

\abstract{
{We propose a novel mechanism for production of baryonic asymmetry in the early Universe. 
The mechanism takes advantage of the strong first order phase transition that produces runaway 
bubbles in the hidden sector that propagate almost without friction with ultra-relativistic velocities. Collisions of such bubbles 
can non-thermally produce heavy particles that further decay out-of-equilibrium into the SM and produce the observed 
baryonic asymmetry. This process can proceed at the very low temperatures, providing 
a new mechanism of post-sphaleron baryogenesis.  
In this paper we present a fully calculable model which produces the baryonic asymmetry along 
these lines as well as evades all the existing cosmological
constraints. 
We emphasize that the Gravitational Waves signal from the
first order phase transition is completely generic and can potentially be detected by the future eLISA interferometer. 
We also discuss other potential signals, which are more model dependent, and point out the unresolved 
theoretical questions related to our proposal.  }
}

\begin{document}
\maketitle

\section{Introduction}
Theories of baryogenesis (for a review see~\cite{Riotto:1998bt, Riotto:1999yt}) aim at explaining the tiny difference
between the number density of baryons and antibaryons, about $10^{-10}$  (in units of the entropy density) we observe in our universe.  Until now, many mechanisms for the
generation of the baryon asymmetry have been proposed. Grand Unified Theories  unify the strong and the electroweak
interactions and predict both baryon and lepton number violation at the tree level. They may be considered 
perfect candidates for a theory of baryogenesis. The out-of-equilibrium decay
of superheavy particles can explain the observed baryon asymmetry, even though there
remain problems strictly related to the dynamics of reheating after inflation and to the fact that the baryon asymmetry generated 
by the decays may be erased by the nonperturbative sphaleron configurations present already in the Standard Model (SM). 

This latter problem is in fact made a virtue in the  leptogenesis mechanism where the cosmic baryon asymmetry 
originates from an initial lepton asymmetry generated in the decays of heavy sterile neutrinos in the early universe and 
then converted into baryon asymmetry by the sphalerons themselves
(for a review see~\cite{Fong:2013wr}). 

If one wishes to resort to much lower energy scales, one should consider the 
electroweak baryogenesis (for reviews see~\cite{Morrissey:2012db,Konstandin:2013caa}), 
where the  baryon number violation takes place during first-order phase transitions
happening at temperatures of the order of the electroweak scale and the baryon number is preserved 
against the SM sphalerons if the
first-order transition is sufficiently strong.  Of course in the SM as it stands with $m_h \approx 125$~GeV there is
no electroweak phase transition (EWPT) and instead one finds a smooth 
crossover~\cite{Gurtler:1997hr,Laine:1998jb,Csikor:1998eu,Aoki:1999fi}. 
Therefore in order to have a viable electroweak (EW)  baryogenesis, 
one should necessarily augment the SM with new degrees of freedom that strongly couple to the Higgs. Moreover, 
new CPV sources at the EW scale are needed.   
These features make electroweak baryogenesis scenarios  attractive because they  could be
tested or ruled out at the LHC and future colliders. However, current LHC results and, in particular, the Higgs
precision data already corner the parameter space of the EW baryogenesis~\cite{Cohen:2012zza,Katz:2014bha} 
and even more gains are 
expected at HL LHC and in future colliders FCC-ee and FCC-pp  (see~\cite{Contino:2016spe} for a recent review of the FCC-pp 
potential reach).  

Can one generate a baryon asymmetry  at temperatures lower than about 100~GeV, when the baryon number 
violation provided by the
SM sphalerons is switched off? 
The three required ingredients for baryogenesis are
of baryon number violation, C and CP violation and out-of-equilibrium 
dynamics~\cite{Sakharov:1967dj}. It is not easy to generate the
baryon asymmetry in a universe that reheats after inflation  to a low
temperature because the first and third ingredients are
hard to come by: it is difficult to introduce baryon
number violation at low temperatures without contradicting
laboratory bounds on baryon number  violation, and the universe
is expanding so slowly at low temperatures that it
is very close to equilibrium. However, several examples of such mechanisms exists, noticeably in the context of baryon-number
violating SUSY at the EW scale. In Ref~\cite{Dimopoulos:1987rk} such a scenario demands a notoriously low reheating
scale of order of dozens of MeV. This unappealing feature can be potentially avoided in a scenario of~\cite{Cline:1990bw}
if very heavy gravitinos are assumed.\footnote{This first of these scenarios is now heavily disfavored by the LHC 
searches for the RPV SUSY and displaced 
vertices~\cite{Liu:2015bma}. } 

Another interesting example of low-energy baryogenesis, that we will later return to, 
was introduced in the context of ``darkogenesis"~\cite{Shelton:2010ta}. In the latter scenario the asymmetry is 
first produced in the hidden sector, which harbors new particles with the baryon number, and later on shared with the visible 
sector. The asymmetry in the hidden sector is produced due to the strong 1st order phase transition (PT), which provides 
low-temperature departure from equilibrium. 

In this paper we 
propose a testable mechanism that overcomes these difficulties and 
naturally produces the baryon asymmetry at the low temperatures, when the sphalerons are not active anymore. 
The scenario that we 
propose crucially relies on the existence of a hidden valley that is endowed with a very non-trivial thermal 
dynamics and is capable of producing heavy particles out of equilibrium, which further decay to the SM via
interactions that violate both the baryon number and the CP.   The hidden valleys as beyond the SM (BSM) particle sector 
with potentially low masses, significantly below the EW scale, but  with weak couplings to the SM via 
higher-dimensional operators (suppressed, say, by the TeV scale) have been proposed in 
Refs.~\cite{Strassler:2006im,Strassler:2006ri,Strassler:2006qa}. 
The motivation 
for this kind of physics was dominantly signature-driven, because production of these new particles promised rare 
but spectacular events at LHC. Nonetheless, it was soon understood that the hidden-valleys scenarios can be highly 
theoretically motivated, for example, being a necessary ingredient of various scenarios, addressing the naturalness problem
of the SM~\cite{Craig:2015pha,Curtin:2015fna,Curtin:2015bka}. 

In this paper we show that the hidden valleys can also be closely related to the problem of the baryogenesis. The basic 
idea of our mechanism is that the strong 1st order PT in the hidden sector can produce non-thermally particles, 
which are themselves much heavier than the temperature of the PT, closely following the old idea of Ref.~\cite{Watkins:1991zt}.
This can happen only in very strong 1st order PT, when the friction of surrounding plasma is not big enough to stop the bubble
walls  of the broken phase from propagating with the speed on light. Collisions of these ultra-relativistic  bubble walls 
can potentially produce particles much heavier than the temperature of the PT itself. 
The decays of these heavy particles, however, dominantly proceed into the SM rather than the hidden sector due 
to the accidental symmetries of the model. The observed baryon asymmetry is produced in these decays. 
Thus an appealing and unique feature of this scenario is that one should neither necessarily reheat the Universe to the temperature 
of the decaying particle, nor produce the decaying particles during the inflation. Reheating to the temperatures above the temperature 
of the hidden PT should be sufficient to trigger the mechanism. In this paper we will both discuss the mechanism 
in detail and introduce a particular model, which is fully calculable and perturbative, satisfies all the existing cosmological
constraints and illustrates all the features of the mechanism, being its existence proof.\footnote{It is not the first 
time when the runaway bubbles are invoked in the model-building 
to produce the Universe baryonic asymmetry. Refs~\cite{Konstandin:2011dr,Konstandin:2011ds}  take 
advantage of the reheating process which that follows the runaway bubbles collisions to produce the baryon 
asymmetry at the EW scale. Ref.~\cite{Servant:2014bla} advocates low-scale baryogenesis from the delayed EWPT
with the runaway bubbles. In spite of these similarities, as we will see, our proposal is quite different. }
 
Of course, because one can probably construct lots of pretty different models along the lines of our proposal, potentially 
including the models that take advantage of non-perturbative dynamics, it is hard to talk about generic 
signatures of this scenario, for example any kind of guaranteed signature for the LHC or future colliders (although 
particular 
realizations might have interesting predictions both for the LHC and the di-nucleon decay experiments). 
However, one signature 
of this scenario is completely generic and virtually inevitable: the gravitational waves. The strong first order PT which must 
proceed in the hidden sector necessarily produces gravitational waves. We will show in our paper, that these gravitational 
waves are of right intensity and frequency to be observed or excluded by eLISA experiment, providing a unique 
opportunity for this future interferometer. 

Our paper is  structured as follow. In the next section we discuss the basic mechanism that we propose in details, discussing 
qualitatively the conditions for the runaway bubbles during the PT and why one might expect heavy particle production
from these collisions. In Sec.~\ref{sec:su2model} we present an existence proof to the mechanism, described in 
Sec.~\ref{sec:mechanism}. We show that this particular model can satisfy all the cosmological constraints, identify 
particular spots in parameter space where phase transition with the runaway bubbles occur and estimate possible 
baryonic abundance. We show that the baryonic abundance produced can be easily be in agreement 
with the observed one. In Sec.~\ref{sec:signatures} we discuss the experimental signature with a special emphasis on the 
primordial gravitational waves and the prospects of eLISA. Finally, in the last section we conclude and discuss some
open questions and  possible
future directions.    
      
\section{The Mechanism Description}
\label{sec:mechanism}
The mechanism that we propose relies on an assumption, that  heavy
particles, $\psi$, which decay out of equilibrium via CPV and baryon-number violating (BNV) or 
lepton-number violating (LNV)
operators, 
are produced {\it non-thermally} at temperatures much lower than their masses.  This will allow us to produce these
particles abundantly even if the Universe is reheated to the
temperature, which is significantly lower than the mass of the
particle $\psi$, or, alternatively if the $\psi$ is annihilated too
efficiently.

If the decays of the particle $\psi$ proceed at high temperatures, above the 
temperature of the EWPT (which, in the absence of the new physics, is simply a crossover),
one can take advantage of, for instance,  LNV (or of a production of any   other charge asymmetry as long as is not 
orthogonal to the baryon number),  as long as this  does not contradict the 
low-energy experimental test. 
If the asymmetry is formed in the leptonic sector, it 
will efficiently moved to the baryonic sector by the SM sphalerons. 
On the other hand, we will be in particular 
interested in the very \emph{low temperature baryogenesis}, both because this 
region of the parameter space is less explored and because this is where we can expect interesting 
gravity wave signature in the detection range of eLISA. If these decays proceed after the EWPT, 
the SM sphalerons are not active anymore and the decay must proceed via the BNV operator. This 
is why we will mostly concentrate on the BNV decays, since they can generically work at any temperature
scale, however we should bare in mind that for the high-temperature decays LNV is also a viable option. 
   
On top of this, the decays of the $\psi$ must proceed via CPV operators in order to 
produce the desired baryonic asymmetry. We will show that in particular realizations that we have 
in mind this CPV can be very efficiently hidden from the low-energy probes like electron and neutron 
EDMs even if the baryogenesis scale is very low. 
Similar to the common lore of the leptogenesis and the
decays of the ``WIMP-like particles'', the sector of the new fermions
should have at least two independent CPV phases. 

Many mechanisms are known to produce stable or metastable particles non-thermally. Usually they rely on decays of 
another heavy particles out of equilibrium. However, here we are interested here in low temperature scenario 
with potentially very low reheating temperature. A non-thermal particle production mechanism, which can efficiently
produce the particles much heavier than the equilibrium temperature,  we will take advantage of is 
the runaway bubble collisions in the strong first order cosmological phase transition.  This mechanism has been 
discovered lots of time ago~\cite{Hawking:1982ga,Watkins:1991zt} and was mostly discussed in the context of the 
SM and the potentially first order EWPT. 

The non-thermal low-temperature production of the heavy particles
crucially demands a strong first-order cosmological PT. Note, that this requirement is very different from the 
usual EW baryogenesis requirement of the strong first order PT during the electroweak 
symmetry breaking (EWSB). The EW baryogenesis requires
quasi-stationary bubbles of the broken phase, namely that the bubbles do not propagate ``too fast" through the 
plasma~\cite{Nelson:1991ab,Joyce:1994zn,Joyce:1994zt}. Practically for successful EWBG one  
demands that $v_w$ is smaller than the speed of sound $\sim 0.3$, while the baryon asymmetry is almost 
$v_w$-independent for the values of $v_w$ between $10^{-3}$ and the speed of sound (see Ref.~\cite{Bodeker:2004ws} and 
review~\cite{Konstandin:2013caa} for detailed discussions).\footnote{This requirement is needed 
because the excitations in plasma must diffuse efficiently in front of the bubble wall. See also~\cite{Kozaczuk:2015owa} 
for a related discussion.} In the our case
we do not desire to produce an \emph{asymmetry} via  the phase transition, but rather  \emph{an abundant population 
of heavy states}. Therefore we will require bubbles which propagate almost with the speed of light through the 
plasma, namely $\gamma \gg 1$, with the regular definitions 
\beq\label{eq:gamma}
\gamma \equiv \left( 1 - v^2_{w} \right)^{-1/2}
\eeq

It can be intuitively explained why do we expect the collisions of ultra-relativistic bubble walls to produce particles
that might be much heavier than the temperature of the PT. In the case of the steady bubbles, the bubble wall 
carries a typical momentum $p \sim T$, therefore their collisions produce only particles with masses
of order $m \lesssim T$, which further promptly thermalize in the primordial plasma. In the case of the runaway bubbles 
the incoming momentum is in fact much bigger, $p \sim \gamma T$. This in principle allows non-negligible production 
of particles of order $\gamma T$, which is much heavier than the temperature of the surrounding plasma 
if $\gamma \gg 1$~\cite{Watkins:1991zt}. 
Of course in oder for the mechanism to work, one should  make sure that the particles that are formed 
do not thermalized in the surrounding plasma and stay out-of-equilibrium.\footnote{Although the collision of the 
ultra-relativistic bubbles is a highly non-equilibrium process, in the proposed 
mechanism it is responsible for production of heavy particles which do not carry any asymmetry by themselves. }  

A condition for sustaining the runaway bubbles in certain cosmological phase transitions 
 was nicely formulated in Ref~\cite{Bodeker:2009qy}.  This reference 
shows, that in a gauge symmetry breaking phase transition the pressure on the interface between the phases 
for $\gamma \gg 1$ can be found by replacing the standard thermal potential by a mean  field thermal potential
$V_{MF}$.
The latter can be approximated as 
\beq
V_{MF} = \frac{T^2}{24} \sum_a m_a^2(\phi) 
\eeq  
when the sum runs over all the particles which are in thermal equilibrium during the phase transition. 
This in turn sets a simple self-consistency criterion for sustaining the runaway bubbles
if they beforehand reach the ultra-relativistic velocity: if at the nucleation 
temperature one finds that the mean field-approximated potential at the symmetry breaking minimum point (as 
determined by the full thermal potential!)
has lower energy than the symmetry preserving local minimum,
the bubbles are expected to be runaway.\footnote{Another criterion for 
the runaway bubbles, based on the calculation of~\cite{Bodeker:2009qy}, 
was introduced in Ref.~\cite{Espinosa:2010hh}. 
We find, in full agreement with~\cite{Chala:2016ykx}, that this criterion is stronger than the 
B\"odeker-Moore criterion, although the models that we consider differ significantly from those, considered in~\cite{Chala:2016ykx}. 
Given the current  theoretical uncertainties in the treatment of the runaway bubbles, we are going to adopt the 
B\"odeker- Moore criterion which is the most conservative.}
 
\begin{figure}
\centering
\includegraphics[width = .49\textwidth]{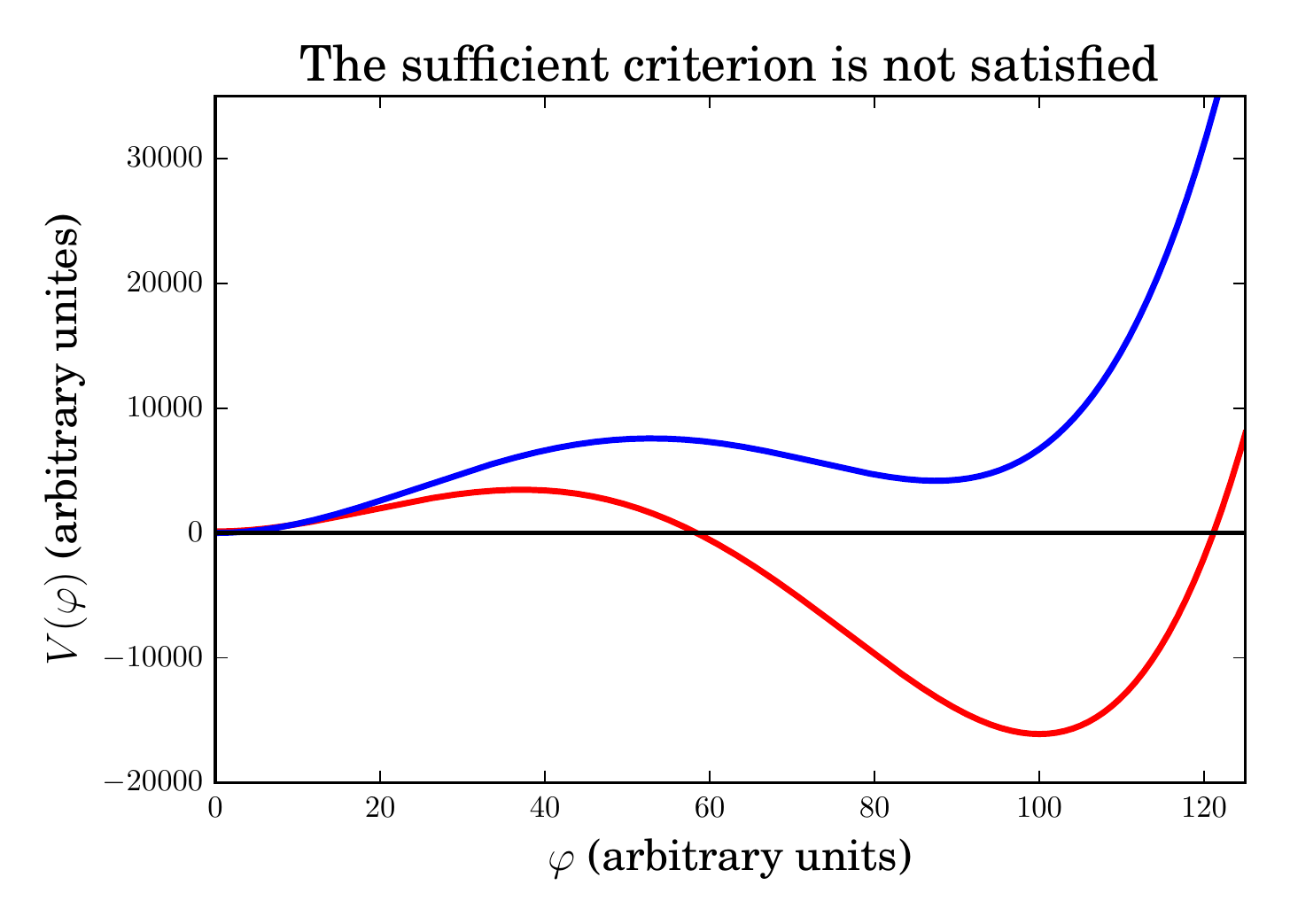}
\includegraphics[width = .49\textwidth]{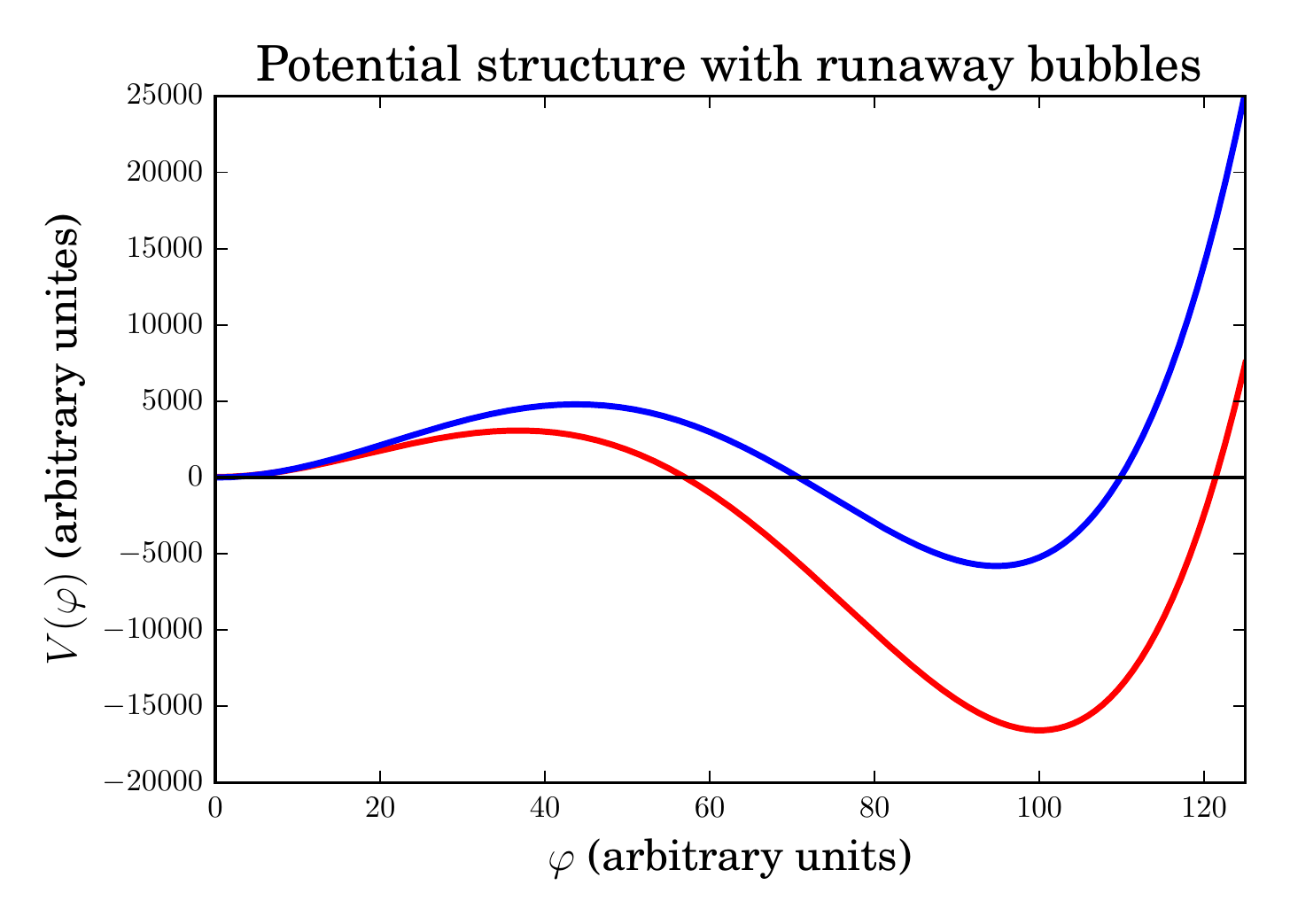}
\caption{Runaway bubbles in potentials which satisfy B\"odeker-Moore criterion. The red line stands for the full one-loop
thermal potential, while the blue line stands for the mean free field approximation. If the potentials look like on the right 
panel, the bubbles during the first order phase transition clearly have runaway behavior. The picture on the left panel 
does not satisfy the criterion: although the local symmetry breaking minimum exists both in the full potential and 
in the mean free field approximation, the vacuum energy of the symmetry breaking minimum  in the mean  
field approximation is bigger than the vacuum energy of the symmetry preserving phase. }
\label{fig:BodekerMoore}
\end{figure}

We will further use this criterion to establish the parameter space of the 
model.
We illustrate this criterion on Fig.~\ref{fig:BodekerMoore}. 

From the very formulation of 
this criterion it is clear that we will further have to focus on the calculable models of  the symmetry breaking PT. 
This does not mean that our mechanism cannot work, for example, in systems with confinement PT. However, since these systems 
are inherently uncalculable and the criterion that would be analogous to the B\"odeker-Moore criterion is not known for these systems, 
we cannot make a firm 
point that such a system produces the runaway bubbles. We believe that some of these 
systems  do, but we cannot show this with our tools and leave this as an open question. It is also worth noticing that
 because our model will rely 
on the higgs-like phase transition, it will not be natural in the same sense as the SM is unnatural. Since we would 
like to present here an existence proof, we will not try to solve this problem here. However, it can be relatively 
easily rendered natural either via supersymmetrization of the hidden sector or via taking advantage of the confinement PT
as we have previously discussed. 

Notice, that from the point of view of hidden valley thermal behavior our mechanism shares some similarities with the idea of the 
``darkogenesis"~\cite{Shelton:2010ta}: both demand strong order phase transition in the hidden sector at temperatures that  are smaller
than the temperature of the EWPT. However, the ``darkogenesis" demands stationary bubbles, because the asymmetry 
is produced in during the dark PT itself, and the dark sector must be augmented with the sufficient sources of the CPV. Our mechanism 
takes advantage of the different regime of this kind of models, with the runaway bubbles and demands no CPV sources in the dark sector
itself. 

Let us now concentrate on the last part of our mechanism, namely the heavy particle that we have introduced. In order 
to be produced abundantly, they should be coupled sufficiently strong to the particles of the hidden sector that undergoes 
the strong first order PT. On the other hand, they should not decay to the particles of that sector. We will concentrate on the models, 
in which these heavy particles possess an approximate $\ztwo$ symmetry, which is exact with respect to the particles of the 
hidden sector and broken by a small coupling to the SM matter to allow out-of-equilibrium decays. Therefore, these heavy 
particles can be either in singlet or adjoint representations of the SM. For simplification purposes we will assume them to the 
the SM singlet fermions. The models with the SM adjoints can be potentially much more challenging, because they can imply 
other particles with the SM charges in the hidden sector.    

From the structural point of view the sector of the heavy decaying particles very closely resembles the analogous structure 
in the baryogenesis for weakly interacting massive
particles~\cite{Cui:2012jh}, 
which can by itself viewed as an incarnation of 
``WIMPy baryogeneis"~\cite{Cui:2011ab}. The crucial difference between our scenario and these works, 
is that by assuming a light hidden 
valley, we can make the entire mechanism operate at much lower temperature scales that the standard WIMP-inspired
scenario.  Moreover, because we can produce even at low scales much bigger abundances of the heavy decaying particles 
that one expects from the thermal scenarios, we will be able to get the observed baryonic asymmetry with much smaller 
values of the CPV, than one needs in the ``WIMPy" scenarios.

\section{A Model}
\label{sec:su2model}
Let us now describe a model that can serve as an existence proof to the mechanism, described in Sec.~\ref{sec:mechanism}. 
Consider an $SU(2)$ gauge symmetry in the hidden sector with ``dark higgs" $\Phi$ in the fundamental 
of the dark $SU(2)$ and a pair of fundamental fermions  $L_i$ with $i = 1, 2$. We allow a standard
symmetry-breaking potential for $\Phi$
\beq
V = - m^2 |\Phi|^2 + \lambda |\Phi|^4~.
\eeq
We will further use $\Phi$ for the full dark-$SU(2)$ doublet and $\varphi$  for the dark higgs excitation, using
following parametrization
\beq
\Phi = \left(
\begin{array}{cc}
0 ,\ \ &   \frac{f + \varphi}{\sqrt{2}}
\end{array}
\right)
\eeq
and $f$ standing for the VEV d the dark higgs. 

 Let us also introduce a pair of singlet
fermions $e_i$, again with $i = 1,2$. The most generic couplings involving the fermions that we can write down are 
\beq\label{eq:DarkFermionL}
\cL \supset y_{ij} \Phi L_i e_j + m_L \epsilon_{ij} \epsilon^{ab} L_a^i L_b^j + (m_e)_{ij} e^i e^j~. 
\eeq
We imagine that the couplings $y_{ij} \sim \cO(1)$, but non-degenarate, while all the rest of the bare masses, $m_L$ and $m_e$,
are of the same order of magnitude as $y \langle \Phi \rangle $. As previously, $i$  is a flavor index 
and $a$ is an $SU(2)$ index.
After the hidden $SU(2)$ breaking and diagonalization of the fermionic mass matrix one gets 6 Majorana fermion 
states roughly at the 
scale $\sim y \langle \Phi \rangle $. Note that these particles generically will inherit $\cO(1)$ couplings to the dark higgs, however
the lightest fermionic states \emph{cannot decay into the hidden sector} because of the accidental $\mathbb{Z}_2$ symmetry, 
under which they are odd and the rest of the hidden sector is even. Namely they perfectly satisfy the criteria, that we have 
laid out in the previous section: heavy particles with $\cO(1)$ couplings to the dark sector but unable to decay into it. We will later 
couple these particles weakly to the SM particles to ensure their decays via BNV and CPV operators to produce the 
visible sector baryonic asymmetry. 

The fermions that we have described above have to be produced in the bubbles collisions during the dark $SU(2)$ breaking 
phase transition. As we will see later explicitly, this will be possible if we choose following regime
\beq\label{eq:regime}
m_{fermions} \gg m_{gauge-bosons} \gg m_\varphi~.
\eeq
Of course this model of the hidden sector is by no mean novel. It strongly resembles the EW model (without gauging 
the $U(1)_Y$), as well as handful of the ``darkogenesis" models. The trick is that in the regime~\eqref{eq:regime}, 
which has not been studied in detail beforehand, the model exhibits a very non-trivial  thermal behavior that we will 
take advantage of.   

The structure of the rest of this section is as follow. First, in~\ref{subsec:runaway} we establish the 
parameter space for the runaway 
bubbles and discuss the production of the heavy fermions in the collisions. However, because the hidden sector 
might populate very 
low mass ranges, one might worry about undesired relics and late decays in this sector. We address these questions 
of the cosmological
safety in~\ref{subsec:safety}. In the subsection~\ref{subsec:couplings} we discuss the couplings 
of the exotic heavy fermions to the SM and 
the production of asymmetry in decays. 
Finally, in~\ref{subsec:asymmetry} we will return back to the heavy fermions sector, 
define the couplings to 
the SM and calculate the expected baryonic asymmetry that we produce. 

\subsection{Runaway bubbles in the hidden sector}
\label{subsec:runaway}
We concentrate now on the higgs and gauge bosons part of the hidden sector. As we have alluded beforehand, we will 
be interested in regime where the gauge bosons are much heavier than the dark higgs, triggering gauge bosons driven PT. This 
in turn, determines the hierarchy of couplings $g^2 \gg \lambda$, where $g$ is a dark gauge coupling and $\lambda$ is a 
dark quartic. It is well known from the studies of the SM with a very light higgs, that the phase transition in this case 
is strong first order, and driven by the gauge bosons. We further calculate the thermal potential for the dark higgs, which 
in our case becomes
\beq\label{eq:FullThermal}
V_{th} = \frac{g_i T^4 (-1)^{F_i}}{2\pi^2} \int_0^\infty dx \ x^2 \log \left(1 - (-1)^{F_i} \exp\left(
\sqrt{x^2 + \frac{m_i^2(\varphi)}{T^2}} \right) \right)
\eeq 
It is well known that although qualitatively this expression is sufficient, it misses lots of numerical effects that have to do with large
two-loop corrections, non-perturbative effects and further uncertainties that we will latter list in detail. However, since from the 
SM lesson we know that it does give qualitatively correct answer, we will proceed with this expression. Because we are merely 
trying here to give a existence  proof, we will be less worried about numerical factors of $\cO (1) $ that we will be likely to miss 
with our procedure. We also perform ``daisy" resummation~\cite{Fendley:1987ef, Carrington:1991hz} 
 by replacing in the thermal potential
\beq
m^2_a(\varphi) \to m^2 + \Pi_a (T)~,
\eeq
with $\Pi_a(T)$ being the thermal masses squared of the particles, usually of order of the plasma temperature. 

In order to check, if the potential satisfies the B\"odeker-Moore criterion, we compare it at temperature 
$T$ to the mean field potential, which in our case takes a compact form
\beq
V_{th, MF} = \frac{3\, g^2 T^2 h^2}{32}~.
\eeq
Of course, in both cases we add the thermal potential to the tree level potential and the one-loop Coleman-Weinberg 
potential.  

It is worth now explaining why do we originally choose the 
regime~\eqref{eq:regime} and why does the system 
in this regime often satisfy the 
sufficient criterion for the runaway bubbles. 
Ref.~\cite{Bodeker:2009qy} pointed out that runaway bubbles in the 
context of the EWSB is reachable with at least two scalar fields in the higgs sector and can 
happen in singlet-catalyzed 
EWSB. The reason is very straightforward: one usually gets the system with two different vacua from the 
\emph{cubic} higgs term in high-temperature expansion of Eq.~\eqref{eq:FullThermal}. 
Because the mean  
field thermal potential only cares about the leading term in $T^2$, it contains no cubic term in the higgs field, essentially 
ruling out the possibility of getting two-vacuum structure in the mean field approximation with a cubic term
that is not produced already at the zero temperature. In the concrete example of Ref.~\cite{Bodeker:2009qy}
the extra singlet has been used to produce a tree level effective cubic term.  

However, our scenario is qualitatively different from the EWSB. Because we insist on the mass 
hierarchy~\eqref{eq:regime}, our quartic coupling in the higgs sector is much smaller than the gauge coupling $g$. 
In this situation we might discover, that in the $T = 0$ potential the 1-loop Coleman-Weinberg terms are as important, 
as the tree level terms. This is indeed the case when 
\beq\label{eq:lambdag}
\lambda \sim \frac{g^2}{16 \pi^2}
\eeq  
and this is precisely a regime that we are going to further focus on.  For this choice of parameter space, the $T = 0$ 
potential has a much more complicated structure, because it also includes the Coleman-Weinberg term
\beq
V_{CW} = \frac{N_g}{64 \pi^2} m_g^4 (\varphi) \left( \log \frac{m_g(\varphi)^2}{v^2} - \frac56 \right)
\eeq
where $N_g$ is the total number of heavy gauge bosons degrees of freedom, which is 9 in the case of 
our simple $SU(2)$ toy model and $m_g = \frac{g\varphi}{\sqrt{2}}$ is the higgs-dependent mass of the heavy 
gauge bosons. Of course, in order to make sure, that the values of the $T = 0$ higgs mass and its VEV do not move 
from the values
that we have chosen, we add appropriate one-loop counter-terms to the $\lambda$ and $m^2$. 

This is not the first time that it noticed that significant changes  in the structure of  $T=0$ potential 
might have an important impact on the nature of the symmetry breaking. Similarly to our structure, 
it was noticed in~\cite{Espinosa:2007qk} that the changes in the \emph{zero-temperature potential}
can trigger the first order PT in the EWSB. This picture is very different from the standard EW first order PT, 
driven by the thermal loops of the new particles. In the picture of~\cite{Espinosa:2007qk}
the barrier between the symmetry-breaking and the symmetry-preserving vacuum may persist even
 at the zero temperature, 
while the thermal loops play relatively a minor role in shaping the thermal potential (the effect was also discussed 
in detail in~\cite{Curtin:2014jma}).   
In the SM case this is not that easy to achieve because 
for the $m_h \approx 125 $~GeV the quartic of the SM is $\lambda_{SM} \approx 0.12$. If one would like 
to have a Coleman - Weinberg (CW) potential that is numerically comparable to the tree level one, he would need essentially 
non-perturbative couplings of the new particles to the higgs. Ref.~\cite{Espinosa:2007qk} suggests 
overcoming this problem by putting large number of new scalars at the EW scale with $\cO(1)$ couplings 
to the SM higgs. However, it is not clear weather the model, presented by this reference is in fact calculable 
in 't Hooft sense, namely $N \lambda \ll 1$. 

Nevertheless, one can easily envision this scenario without dubiously large couplings 
or high-multiplicity fields in the hidden sector. Moreover, 
the situation emerges naturally in the regime~\eqref{eq:lambdag} where the corrections to the 1-loop CW
potential at $T = 0$ from the gauge bosons are so significant, that they in fact reproduce the picture of the 
zero-temperature produced barrier.    

\begin{figure}
 \centering
 \includegraphics[width = .65\textwidth]{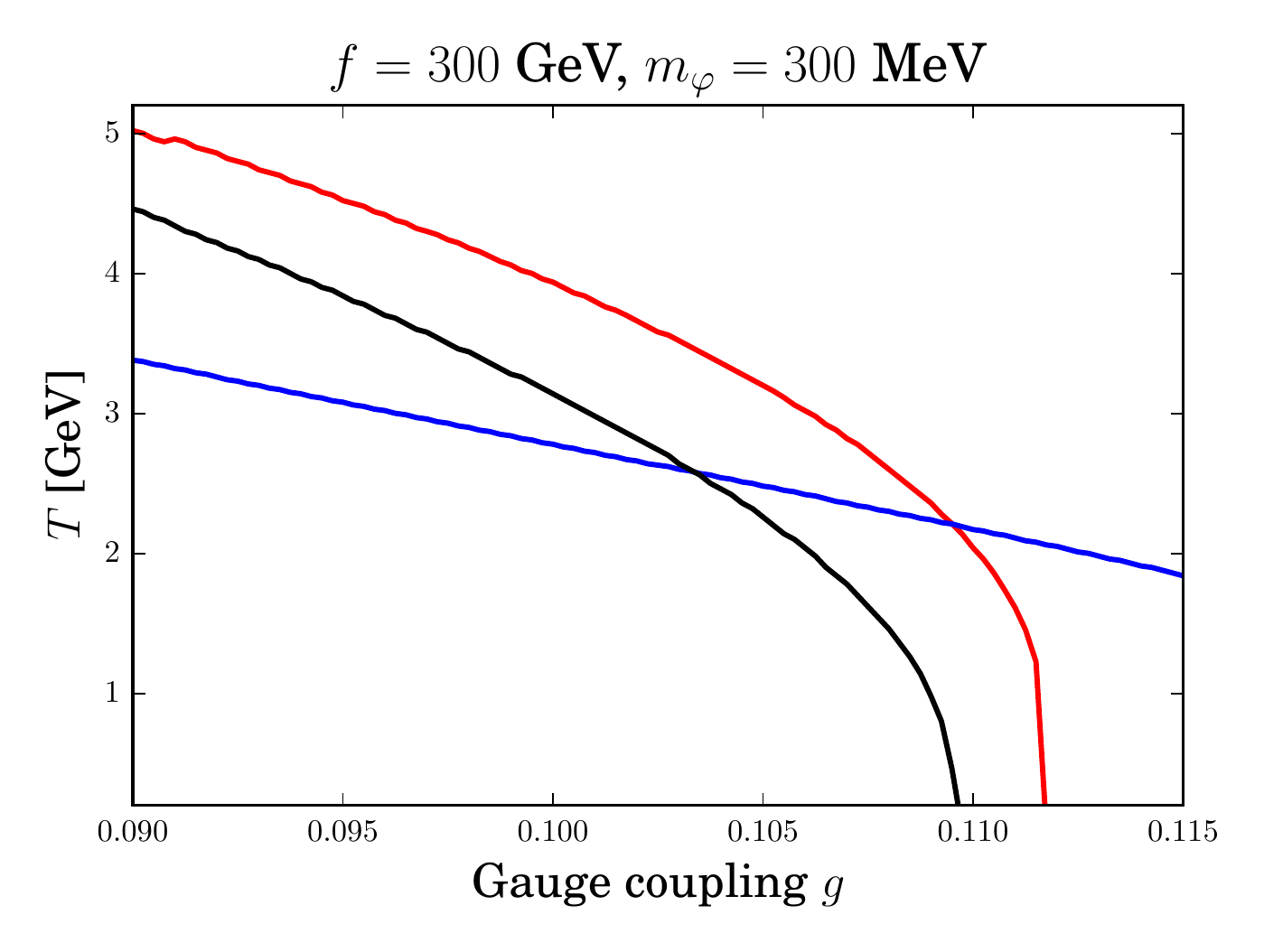}
 \caption{The upper bound on the nucleation temperature (red line) and the  highest temperature at which BM 
 criterion for the runaway  bubbles is satisfied (blue line). Below the black line the barrier between the real vacuum 
 and the false vacuum disappears. In the most pessimistic scenario the relevant parameter space 
 lies to the right of the intersection between the blue and the red curves, while in the most optimistic 
 scenario it is to the right of the intersection between the  black and the blue curves. 
 }
 \label{fig:BMconditions}
 \end{figure}

To check further for the candidate points in the parameter space we proceed as follow. We identify the points 
which satisfy the B\"odeker-Moore criterion for a certain temperature and calculate the upper bound on the 
nucleation temperature for these points. If the $O(3)$ solution has the least action, 
the tunneling probability per unit time per unit volume is given by~\cite{Linde:1977mm,Linde:1981zj}
\beq
\frac{\Gamma}{\nu} = A(T) e^{-S_3/T}
\eeq 
where $S_3$ is the three-dimensional bounce action and $A(T)$ is a prefactor of order $\sim T^4$.  
In order to calculate the bounce action, we should solve for $\phi$ the Euclidian 
equation of motion
\beq\label{eq:ExactWaveFunctionEq}
\frac{d^2\varphi}{dr^2} + \frac{2}{r} \frac{d\varphi}{dr} = \frac{dV(\varphi; T)}{d\varphi}~,
\eeq
which is a boundary value problem 
with the boundary conditions $\varphi(\infty) = 0$ and $\dot \varphi(0) = 0$. One further calculates the 3D 
bounce action using this approximation.

One can make a further approximation by assuming that the thickness of the bubble wall at the moment of its formation 
is much smaller than the radius of the bubble~\cite{Linde:1977mm,Linde:1981zj}. In this case the 3D actions can 
be approximated as 
\beq\label{eq:Linde}
S_3 \approx -\frac{4\pi}{3} r^3 \Delta V + 4 \pi r^2 S_1, \ \ {\rm where } \ \ S_1\equiv \int d\varphi \sqrt{2 V(\varphi; T)} 
\eeq
where $r$ stands for the radius which extremizes the 3D action, $\Delta V$ is the energy difference between the 
false minimum and the global one, and the integration in the surface term $S_1$ runs from zero to the point at which 
$V(\varphi_{end}) = V(\varphi = 0)$. In fact, this approximation is decent if $\Delta V$ is much smaller than the barrier between 
the minima. This corresponds to the \emph{weak first order PT}, which is not our case. Therefore, strictly speaking the 
approximation~\eqref{eq:Linde} never works in our case. However, it had been shown in Ref.~\cite{Dine:1992wr} that 
this approximation always overestimates the temperature of the PT, 
henceforth it always underestimates the strength of the 
PT. The underestimation is in fact as big as $\cO(1)$. 
 
Finally after the we have calculated the 3D action as a function of temperature, the probability for a single bubble to nucleate within 
the horizon volume is order-one when following condition holds~\cite{Moreno:1998bq}: 
\beq
\xi \equiv \int_{T_{nuc}}^\infty \frac{dT}{T} \left( \frac{2 \zeta M_{pl}}{T} \right)^4 e^{ - \frac{S_3(T)}{T}} \sim \cO(1)~,
 \eeq
 where 
 \beq
 \zeta^{-1} \equiv 4\pi \sqrt{\frac{\pi g_*(T)}{45}} ~.
 \eeq

We show  the bound on the nucleation temperature vs the minimal temperature at which the B\"odeker-Moore criterion 
for the runaway bubbles is satisfied for a particular choice of $m_\varphi$ and $f$ on Fig.~\ref{fig:BMconditions}. 
Clearly  we see that 
the upper bound on the nucleation temperature is lower than the highest possible temperature at which the bubbles 
run away for an appropriate choice of the gauge coupling $g$. Note also that because the PT is very strong first order
in the entire range presented 
on the Figure, the bound on the nucleation temperature is expected to overestimate the real nucleation temperature 
vastly. Therefore, it is not unlikely that the entire parameter space, or much bigger portion of it, 
where the minimal PT temperature  does not exceed the bound on the BM criterion ($g \gtrsim 0.103$), is suitable for the 
runaway bubble collisions. Note also, that the bound on the nucleation temperature that we find, is in fact significantly 
bigger than the dark higgs mass, while at least naively we expect it to scale as the higgs mass.      

\subsection{Cosmological safety of the hidden sector}
\label{subsec:safety}
Although the model that we propose in this paper is merely an existence proof to the larger new mechanism, and 
probably should not be taken too seriously for the phenomenological purposes, one should still check that it is not in conflict 
with the basic cosmological observations. Here we show that all the potential problems 
can be resolved in the relevant regions of the parameter space.

The potential problems are twofold: 
\begin{enumerate}
\item The dark higgs is the lightest dark particle without any obvious efficient annihilation channels. We should introduce
new couplings, to make sure that  the higgs decays into the SM faster than within 1~sec to ensure the 
Big Bang Nucleosynthesis (BBN) safety
\item The SU(2) gauge bosons themselves are dark-stable. A-priori this might be a worry. We will show, that in the full 
theory, where we include the fermions and the interactions with the SM, they decay to the SM states faster than within 1~sec. 

\end{enumerate}

The first problem on this list can be addressed in several different ways. The most straightforward one
that we will focus on is simply via introducing a 
tiny higgs-portal coupling to the SM higgs portal:
\beq\label{eq:HiggsPortal}
\cL = \kappa |H|^2 |\Phi|^2 
\eeq 
One should be careful with this kind of coupling because it also induces an undesired mixings between the 
visible and the dark higgs, as well as it induces invisible higgs decays, which are now constrained by the LHC 
data.   Moreover, while such a coupling induces new mass parameters both to the visible and the hidden higgses. We
explicitly check that the induced $T =0 $ mass does not exceed the bare mass, triggering a potential instability 
of the higgses potentials. 

In the mass insertion approximation
the coupling~\eqref{eq:HiggsPortal} induces following decay rates of $\phi$ into the SM fermions
\beq
\Gamma(\varphi \to f \bar f) = \frac{\kappa^2  f^2 m_f^2 m_\varphi}{4 \pi m_h^4} \left(
1- \frac{4 m_f^2}{m_\varphi^2} \right)^{3/2}
\eeq 
while the invisible decay width of the SM higgs into the hidden higgses is 
\beq
\Gamma (h \to \varphi \varphi) = \frac{1}{8 \pi} \frac{\kappa^2 v^2}{m_h} 
\sqrt{1 - \frac{4 m_\varphi^2}{m_h^2}}~. 
\eeq

The invisible higgs width has been looked for both by ATLAS and CMS 
collaborations~\cite{ATLAS-CONF-2013-011,CMS-PAS-HIG-13-018}, reporting exclusion of the SM-like
higgs invisible branching ratios of 65\% and 75\% respectively at 95\% confidence level.  Of course a  proper 
combination of these results is beyond the scope of this work and we will crudely assume for the further 
analysis that the invisible higgs BR should not exceed 50\%, providing us with an upper bound on $\kappa$.

\begin{figure}
\centering
\includegraphics[width = .49\textwidth]{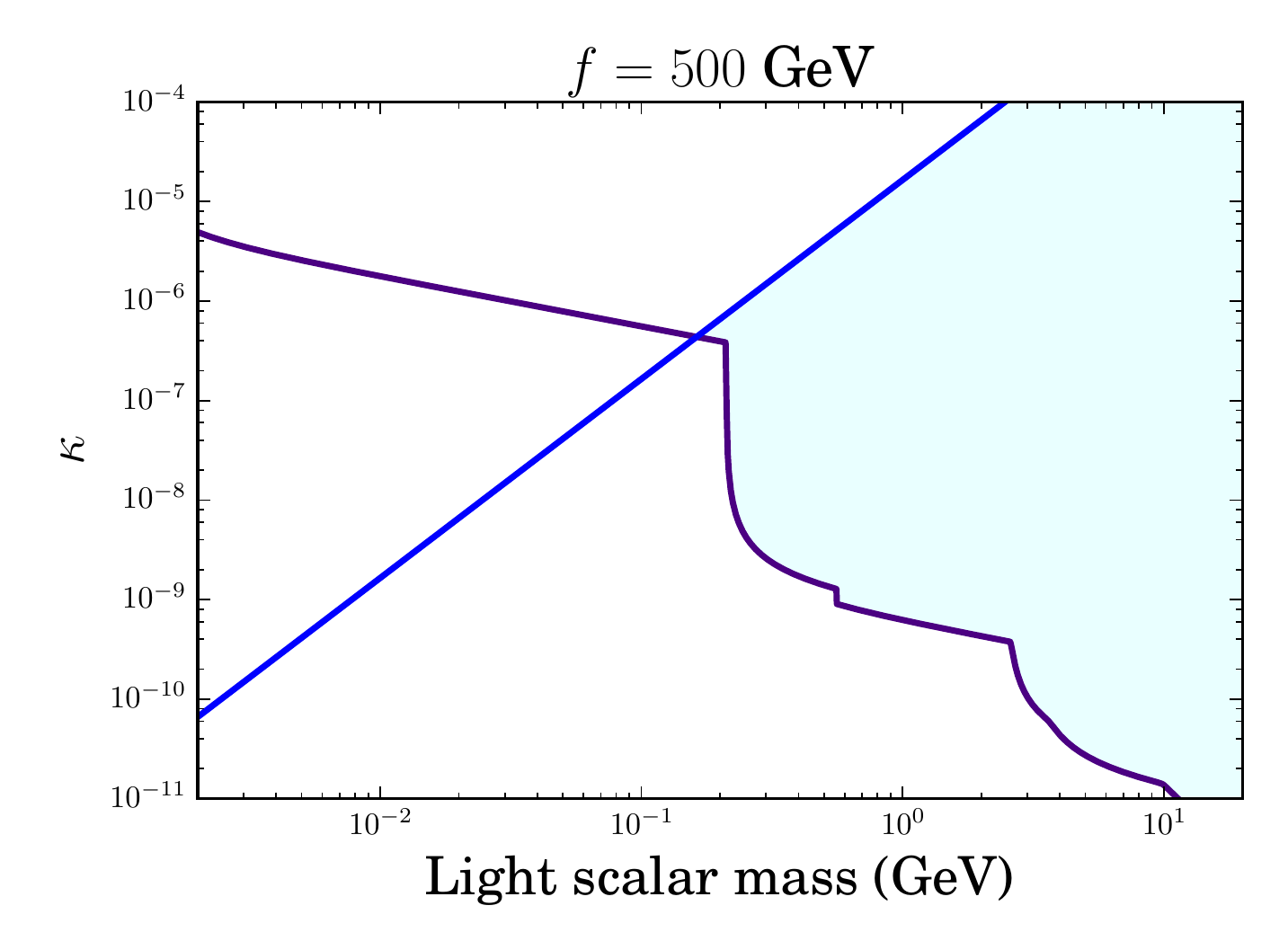}
\includegraphics[width = .49\textwidth]{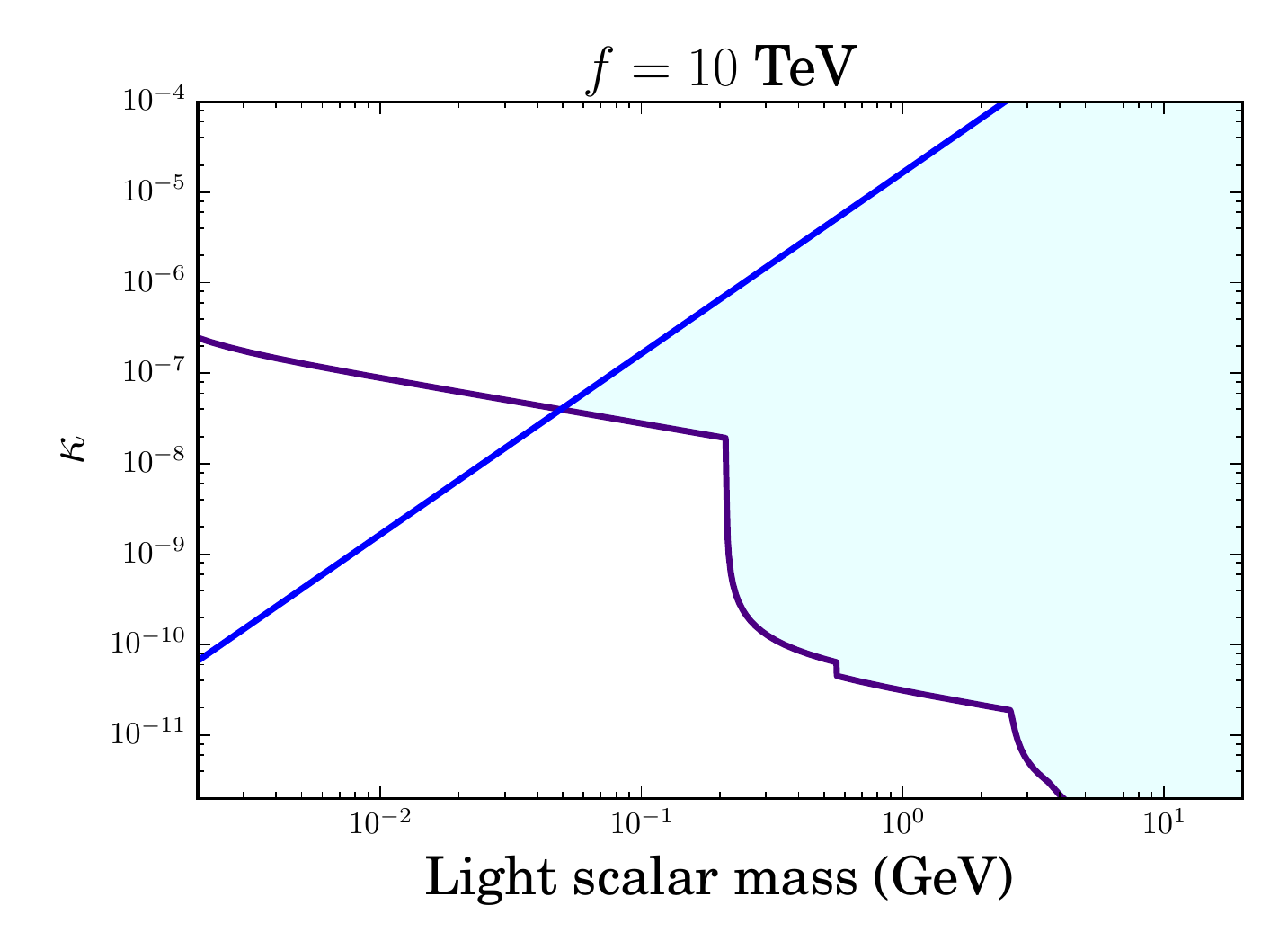}
\caption{Constraints on the value of $\kappa$ in the higgs portal coupling for different values of $f$.
Above the blue line the structure of the hidden sector potential may be altered without appropriate 
fine-tuning due to too large induced dark 
higgs
mass. Below the purple line the decay time of the dark higgs exceeds 1~sec. The shaded region is allowed. The constraints 
from the invisible higgs decays $\kappa \lesssim 0.01$ are completely subdominant and they are not shown on these plots.  }
\label{fig:Hportal}
\end{figure}

The lower bound on $\kappa$ comes from the requirement that $\varphi$ should decay with at 1~sec
in order not to spoil the BBN predictions, or in other words
 $\Gamma(\varphi \to f \bar f) < 6.58\times 10^{-25}$~GeV. Although this constraint looks pretty innocuous 
 at the first glance, it turns out to set a meaningful and  important bound on the light scalar mass. Note that below the mass of 
 1~GeV the scalar is allowed to decay only into the light fermions, suffering both from the small $\kappa $ suppression 
 and from the SM Yukawa suppression.\footnote{Although the decays into di-photons are kinematically allowed, 
their branching ratio (BR)
 never exceeds the the BR into the fermions, as long as the decays into $e^+ e^-$ are kinematically allowed.  The same 
 is true for the decay into the gluons above the $\Lambda_{QCD}$ scale. }  
 On top of this, one gets an upper bound on $\kappa$ from the demand of the stability of the dark higgs potential. 
 We summarize all these constraints for representative choices of $f$ on Fig.~\ref{fig:Hportal}.
 
 As we see, the constraints coming from the higgs invisible width are completely subdominant and the constraints are dominated 
 by considerations of the structure of the dark higgs potential and the fast enough decay of the dark higgs. The parameter space 
 widely opens up around $m_\varphi \sim 100$~MeV.  Here we emphasize that the bound that we get here is only 
 relevant for this particular model, otherwise the scale of the \emph{mechanism, rather than the model} can be as low as
 the BBN scale 1~MeV. 
 
 Another potential cosmological worry is the dark-stability of the massive gauge bosons. However, even in 
 the dark sector itself the accidental ${\mathbb Z}_2$ symmetry which protects the stability of the dark 
 gauge bosons, will be broken by the couplings to the dark fermions. Further, when the interactions with the 
 SM fermions will be taken into account, we will see that these dark massive gauge bosons can potentially decay 
 way faster than 1~sec into the SM fermions and therefore pose no real cosmological threat.  
\subsection{Couplings to the SM}
\label{subsec:couplings} 

After establishing the parameter space of the runaway bubbles PT and setting the scales which do not contradict the 
existing cosmological observations, we are now ready to introduce the heavy fermions sector and later to estimate the 
final baryon asymmetry.  

Let us see how can we couple it to the BNV operators in the 
SM and estimate the baryon asymmetry that they produce. Clearly, we cannot couple $L_i$ particles 
to the SM, because such a coupling would necessarily be non-gauge invariant under the hidden $SU(2)$. 
However, the gauge singlets $e_i$ come here to our rescue. The most generic couplings to the BNV SM operators 
one can introduce are 
\beq\label{eq:FermionInteractions}
\cL \supset \frac{1}{\Lambda^2} \left( \lambda_{\alpha ijk}e_\alpha u^c_i d^c_j d^c_k +
\eta_{\alpha ijk} e_\alpha  Q_i Q_j (d^c_k)^\dagger \right)~.
\eeq
It does not really matter at this stage what is precisely the operator that we further proceed with.
Note, however, that these operators have a slightly different  flavor structure, which will be important for our further discussion 
in Sec.~\ref{sec:signatures}. While the first operator is antisymmetric in the color space, and therefore demand also 
anti-symmetrization over $j$ and $k$ indices in the $\lambda$ coefficients, the second operator is anti-symmetrized
in the color space and in the $SU(2)$ space, allowing non-zero $i = j$ coefficients inside $\eta$.  

We further analyze the decays of the heavy hidden fermions, which eventually produce the baryon asymmetry.
Let us collectively call all the Majorana fermions, that we get from the diagonalization of the fermion matrix $\psi$. 
 For concreteness we concentrate on the couplings $\lambda_{\alpha ijk}$ in Eq.~\eqref{eq:FermionInteractions} and set all the 
coupling $\eta $ to zero.\footnote{Our following discussion has
a handful of similarities to TeV-scale model baryogenesis of Refs.~\cite{Babu:2006wz,Dev:2015uca} because 
of the very similar couplings structure. However, there are important 
differences. First, it will be much simpler, because we neither involve the neutrino masses in our discussion nor invoke the resonant 
processes. Second, more important, we introduce \emph{two generation of Majorana fermions}
 $e_\alpha$, which allows us, similar
to the leptogenesis scenario, to get irremovable phases in the exotic fermions sector, without relying on the tiny Jarlskog
invariance. }

To simplify further the discussion let us UV-complete the four-fermi effective operator~\eqref{eq:FermionInteractions} via 
introducing an exotic colored scalar $\Delta$ with the hypercharge $-2/3$:
\beq\label{eq:Delta}
\cL \supset \lambda'_{\alpha i} \Delta e_\alpha u^c_i + \lambda''_{jk} \Delta^* d^c_j d^c_k + M^2_{\Delta} |\Delta|^2~.
\eeq 
In general, all the coefficients $\lambda'$ and $\lambda''$ are complex. Note also that the fermions $e_\alpha$ are 
\emph{real} due to the Majorana mass term ({\it cf.} Eq.~\eqref{eq:DarkFermionL}) and therefore they cannot 
be re-phased. 

If we forget for a moment about the SM Yukawa couplings, all the imaginary phases of the coupling 
$\lambda''$ can be absorbed into the re-phasing of the RH down-type fermions. If we had been forced to rely only 
on those couplings, the only CPV effects would be proportional to the Jarlskog invariant, forcing our CP-efficiency to be 
as small as $\epsilon_{CP} \lesssim 10^{-8}$. 

\begin{figure}
\centering
\includegraphics[width = .95\textwidth]{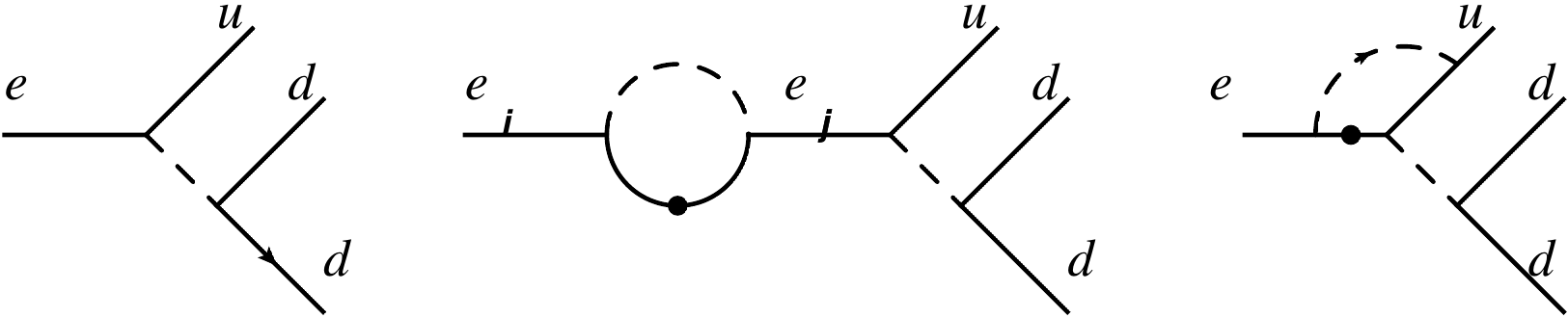}
\caption{CP violating decays of the heavy fermions. In order to produce the asymmetry, one needs the interference
between the (first) tree level diagram and the loop level diagrams, which are sensitive to the irreducible phases of the 
couplings of the scalar $\Delta$. The blobs stand for the mixings between the Majorana particles, that we 
necessary have in the most generic model. These mixings are unavoidable because all the phases in the 
single-generation couplings are removable.  }
\label{fig:HeavyDeacays}
\end{figure}

To avoid this marginal possibility, one can rely on the coefficients $\lambda'$. If we had had only one generation of 
$e_\alpha$, we would have faced the same problem here, because all the imaginary phases could be absorbed   into 
redefinitions of $u_i^c$. However, if we can two generations of the fermions $e_\alpha$ we have three irremovable 
phases, very similar to the common lore of the standard leptogenesis scenario. Also, similar to the standard leptogenesis 
scenario, one gets the CPV violating effect from the interference between the tree level and the one-loop diagram as 
schematically shown on Fig.~\ref{fig:HeavyDeacays}. With the assumption that the phases can be as big order-1 and the 
splittings between various Majorana states are comparable to the masses of the fermions $\psi$, the maximal 
value of the CP-efficiency we can get in a non-resonant perturbative scenario is 
\beq
\epsilon_{CP } \lesssim \frac{1}{8\pi} \left( \frac{m_\psi}{m_\Delta} \right)^2~. 
\eeq      

As we will see in the next subsection, this allows us  lots of freedom with choosing parameters, before 
we even get close to the collider bounds imposed by the LHC. First, we will see that we do not need very high 
value of $\epsilon_{CP}$, and the values as small as $10^{-4} \ldots 10^{-7}$ can be perfectly sufficient in order 
to get an observed relic abundance. This safely allows us to keep the colored scalar $\Delta$ more than an order of magintude 
heavier than the fermions $\psi$. Second in order not to suppress $\epsilon_{CP}$ further, it is sufficient to keep the couplings 
$\lambda'$ to be $\sim \cO(1)$. The couplings $\lambda''$ essentially play no role in contributing to the CP-aymmetry. 
On the other hand the scale $\Lambda$ from the Eq.~\eqref{eq:FermionInteractions} maps onto $\Lambda^2 \sim M_{\Delta}^2 / 
(\lambda' \lambda'') $. Hence by keeping $\lambda''$ small, we can make the scale $\Lambda$ almost arbitrarily 
large without suppressing the CPV effects, easily above the generic bounds of $\sim 100$~TeV, which apply from the 
double-nucleon decay (see Sec~\ref{sec:signatures} for more details). 

\begin{figure}
\centering
\includegraphics[width = 0.7\textwidth]{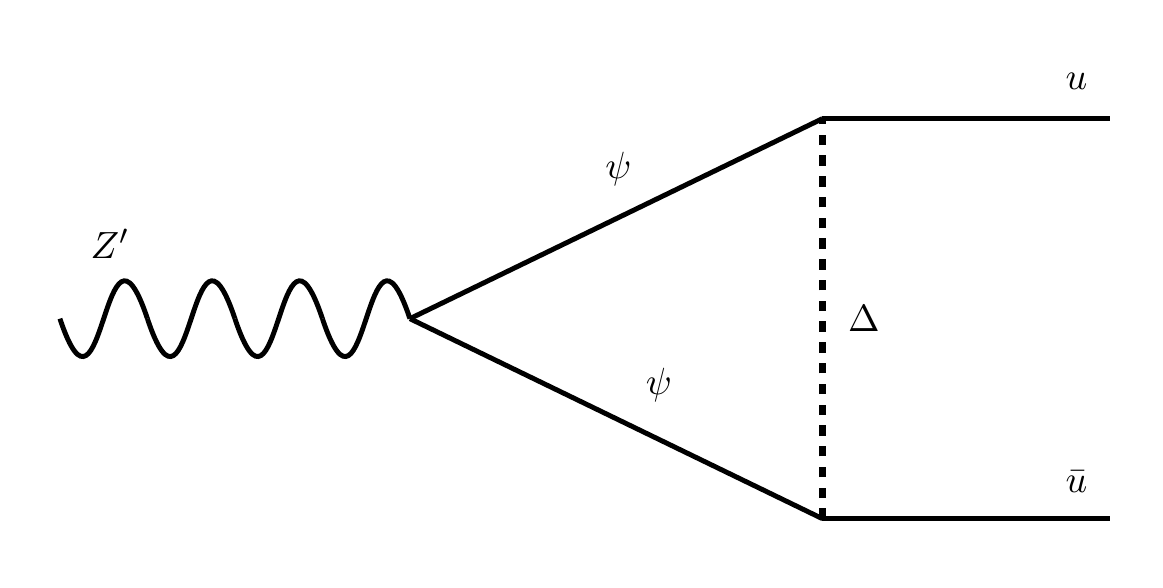}
\caption{Dark gauge boson decay at one-loop, induced by the couplings of the heavy fermions and the scalar 
$\Delta$ to the SM. The coupling to the fermions is gauge coupling suppressed, but assuming $\lambda' \sim1$ there are 
no suppressions in the Yukawa interactions.}
\label{fig:1loopdecay}
\end{figure}

At this point we can return to the dark gauge bosons and briefly comment on their lifetime. We will explicitly 
see that it poses no cosmological worry.  When the effects of the heavy fermions and the scalar $\Delta $
are properly taken into account, we find that the dark gauge bosons can decay to a pair of the SM fermions already
at the one loop level. The decay proceeds via the gauge coupling and the couplings $\lambda'_{\alpha i}$ in 
Eq.~\eqref{eq:Delta} (see Fig.~\ref{fig:1loopdecay}).  The decay rate is parametrically 
\beq
\Gamma(W' \to u \bar u) \sim \frac{1}{8 \pi} \left( \frac{g {\lambda'}^2}{16 \pi^2} \right)^2 
\frac{m_{W'}^5}{m_\Delta^4}  
\eeq
Note, that although we assume that the gauge coupling is small, say $g \sim 10^{-2}$, the coupling 
$\lambda'$ is not necessarily small, and it can easily be $\cO(1)$, while the scale $\Lambda$ can be 
easily small because  $\lambda''$ is small. For representative values $m_\Delta \sim 1$~TeV, 
$m_{W'} \sim 10~\gev$ and $g \sim 10^{-2}$ we get that the decay width is approximately 
$10^{-16}~\gev$, clearly faster than 1~sec, posing no cosmological problem.  

\subsection{Production of heavy matter in bubble collisions and estimation of the asymmetry}
\label{subsec:asymmetry}

Now we can finally calculate the abundance of the heavy fermions produced in the bubble collisions. Because we expect 
dominantly post-sphaleron decays and no important sources of the washout, we expect final produced 
baryon asymmetry to be $\Omega_B h^2\approx \epsilon_{CP} \times \Omega_{\psi} h^2$. 

The abundance of heavy particles produced in runaway bubble have been explicitly calculated in Ref.~\cite{Falkowski:2012fb}
with assumptions of either purely elastic or purely inelastic collisions.\footnote{In this subsection, when we discuss the elastic vs. 
inelastic collisions of the DM, we mean the collisions of the bubble walls rather than the particles. }  
In general, the energy produced in these collisions 
per unit area  in the form of particle $a$ is given by 
{\small
\beq \label{eq:EnergyDensity}
\frac{{\cE }}{A} = \frac{1}{4 \pi^2}\int_{\chi_{min}}^\infty d\chi f(\chi) \sqrt{\chi} \int d\Pi_a \left|
\bar \cM (\varphi \to \alpha) \right|^2 =
 \frac{1}{2 \pi^2} \int_{\chi_{min}}^\infty d\chi f(\chi) \sqrt{\chi}\  \im \tilde \Gamma^{(2)} (\chi)~, 
\eeq
}
where the function $f(\chi)$ carries the information about the details of the bubble 
collisions and the efficiency of production of various particles 
in the runaway bubble collisions depending on the nature of the collision (elastic or inelastic respectively).  $d\Pi_{\alpha}$ 
is a Lorentz-invariant phase space and $\tilde \Gamma^{(2)}$ is a two-point 1PI Green function in the momentum space.  
$\left| \cM(\varphi \to \alpha) \right|^2$ is  spin-averaged squared amplitude of the dark higgs decay into a set of particles~$\alpha$.

In expression~\eqref{eq:EnergyDensity} the integration runs over all the available invariant masses squared of the collisions, 
parametrized by variable $\chi \equiv \omega^2 - \vec p^2 $. $\chi_{min}$ stands for the minimal available invariant 
mass at which the production of the final state $\alpha$ is kinematically possible and it is equal to the square of the sum 
of all the particles masses in the final state $\alpha$, namely $\chi_{min} = \left( \sum_{{\rm state}\ \alpha} M_i \right)^2$.   

We will further analyze two different cases, differentiating the elastic bubbles collisions from the inelastic ones.  
The explicit expression for the function $f(\chi)$ in the elastic case, obtained in Ref.~\cite{Falkowski:2012fb},
is 
\beq\label{eq:elasticF}
f (\chi)_{EL} = \frac{16 f^2 \log \left( \frac{2 (\ratio)^2 - \chi + 2 \ratio \sqrt{(\ratio)^2 - \chi}}{\chi} \right)}{\chi^2} \Theta\left( \left( \ratio
\right)^2 - 
\chi \right)
\eeq
where $\gamma_w$ is defined in Eq.~\eqref{eq:gamma}, $\Theta $ stands for the step function and $l_w$ is the 
thickness of the bubble wall. Practically one usually gets $l_W \sim \cO(10) / T_{PT} $. The above written approximation
assumes the thin wall approximation. The reader can find the full expression, without a thin wall approximation in 
Ref.~\cite{Falkowski:2012fb},  however we made an explicit numerical check and found 
 that~\eqref{eq:elasticF} is sufficient for our purposes and does not introduce 
intolerable numerical imprecisions. The analogous expression for the purely inelastic collisions is 
\beq\label{eq:inelasticF}
f(\chi)_{IN} = 4 f^2 m_\varphi^4 \frac{\log\left( \frac{2 (\ratio)^2 + \chi + 2 \ratio \sqrt{(\ratio)^2 + \chi}}{\chi} \right)}{\chi^2 
\left( (\chi-m_\varphi^2)^2 + m_\varphi^6 (\ratio)^{-2} \right)}
\eeq 
Parenthetically we note that this expression is also an approximation, although it is also adequate for us in the entire 
parameter space.  

The work of Falkowski and No have analyzed the efficiency of the elastic and inelastic collisions in the case of the EWPT, 
namely where one naturally has $T_{PT} \sim m_h \sim v$. They concluded that there is no efficient production above the 
scale $m_h$ (already in striking disagreement with the original conjecture of Watkins and Widrow~\cite{Watkins:1991zt}, 
which talks about efficient production of particles as heavy as $m \sim \gamma T$), 
while the situation is more subtle with the elastic collisions. The paper by Falkowski and No emphasizes
 that while scalars heavier than the 
higgs mass cannot be produced in the 
EW theory runaway PT, it is possible that heavy exotic gauge bosons and fermions can still be produced abundantly. 
While we agree with the statement on the scalars, we find that such a production in the SM is impossible 
also for the gauge bosons and the exotic fermions. The problem is that in order to ensure sufficient production of the 
exotic particles above $T_{PT}$, Ref.~\cite{Falkowski:2012fb} assumed the tree level couplings of the exotic fermions 
to the Higgs $\cL \supset \lambda H f \bar f$ and the tree level coupling of the higgs to the exotic vector  bosons
$\cL \supset \lambda _V h M_V V_\mu V^\mu$, with $H$ standing for the SM higgs doublet and $h$ is standing for the 
physics higgs boson.  
In the former case it is impossible to talk about production of particles 
much heavier than the $m_h$ because this would imply non-perturbative Yukawa coupling to the higgs.\footnote{A theoretical
possibility that this exotic fermion can gain its mass partially from the non-higgs sources would imply additional 
sources of the SM $SU(2) \times U(1)$, the possibility which is cornered by the EW precision tests.} In the latter case an 
efficient production of the heavy exotic gauge boson will jeopardize the unitarity of the $VV \to VV$ scattering.

After the short overview of the EWPT, let us return to our case that allows for a separation of scales $m_h,\ f$ and 
$T_{PT}$, which does not happen in the SM. We find that expressions~\eqref{eq:elasticF} and~\eqref{eq:inelasticF}
behave almost identically at $\sqrt{\chi} < m_\varphi$, with an inelastic solution having a characteristic 
resonance at $\sqrt{\chi} \approx m_\varphi$.  However the behavior at higher masses is very different: while the $f(\chi)_{IN}$
drops very fast above the scale $m_\varphi$, rendering the production of heavy particles at $\sqrt{\chi} > m_\varphi$
virtually impossible, the function $f(\chi)_{EL}$ descends much less steeply. We find that in purely elastic collisions one can still 
efficiently produce in the runaway bubble collisions particles as heavy as $f$. Above this scale, both elastic and 
inelastic collisions turn out to be highly inefficient. Therefore, it is clear, that in order for our mechanism to work, we will need 
the collisions to be fairly elastic.

Another worry that we should emphasize here is the possibility of elastic collisions and the possibility of production of particles 
 in the elastic collisions in general. It is entirely possible that in the case of perfect
elasticity the bubbles simply travel  through one another without producing any particles in the collisions.\footnote{We are grateful to Thomas Konstandin for an illuminating discussion on this 
issue.} This is the case, for instance,  of the Sine-Gordon model, as noticed already in \cite{Watkins:1991zt},
 for which the production 
becomes possible only at $\gamma_w^{-1}$ order. Numerical simulations though show that the dynamics of bubble collisions 
is extremely complicated~\cite{Kolb:1997mz}:   in models with a $\varphi^4$ models bubbles collide and indeed  
may reflect each other by also  emitting soft scalar waves. We leave the decisive conclusion on this point to  future 
studies.

\begin{figure}
\centering
\includegraphics[width = .63\textwidth]{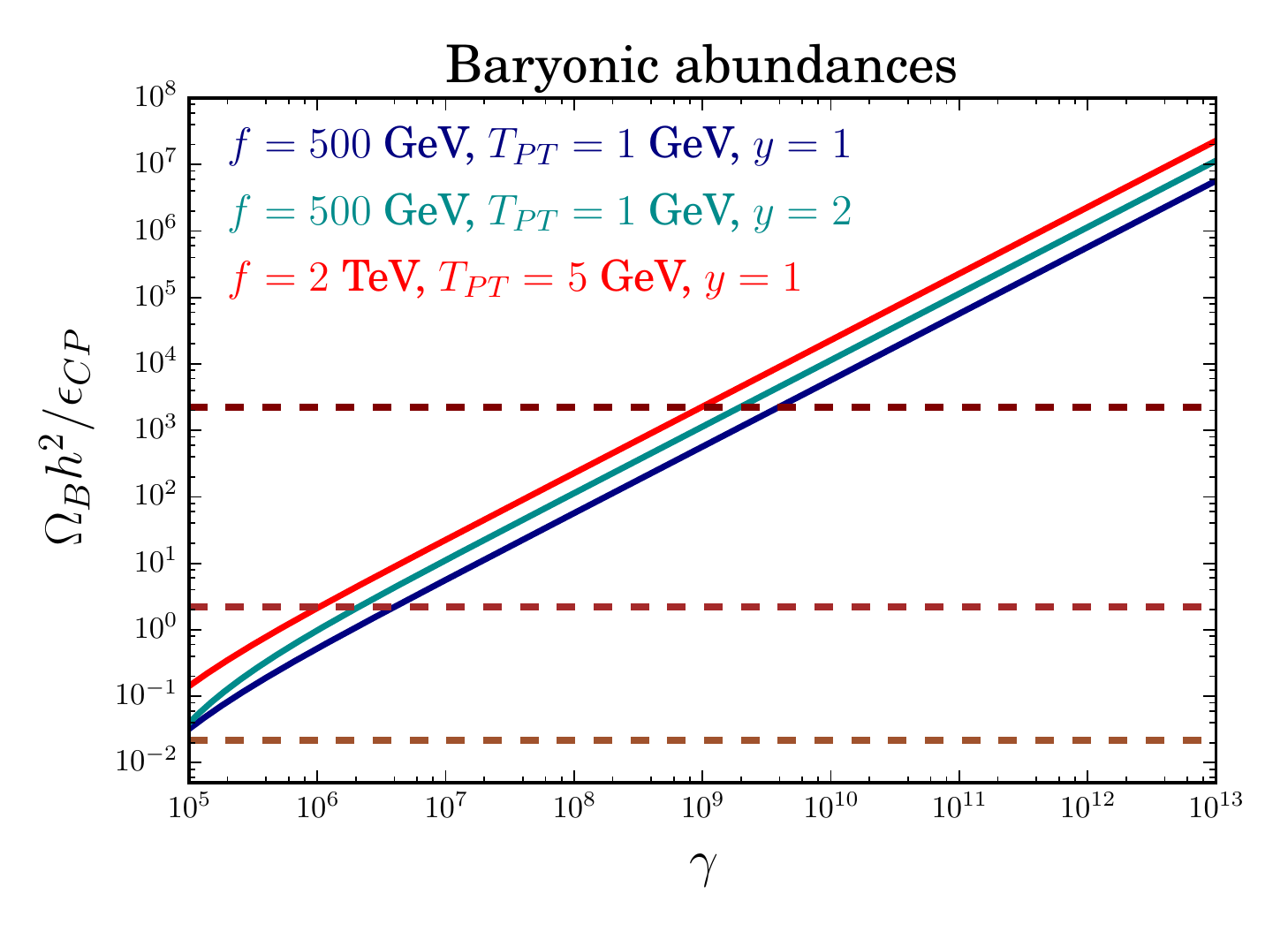}
\caption{Predicted baryonic abundances (reweighed by the CP-efficiency of the process) as a function of $\gamma$. 
The dashed lines bottom-up show the measured $\Omega_B h^2$ (namely requiring  perfect CP-efficiency), 
$100 \times \Omega_B h^2$ and $10^5 \times \Omega_B h^2$ respectively. In all the cases we have assumed $\beta^{-1} = 
10^{-3}\, H^{-1}$. }
\label{fig:abundances}
\end{figure}

Now we can use all these formula and, assuming the elastic collisions, estimate the total abundance of the baryonic matter 
in the Universe produced by our mechanism. The relevant two-point 1PI Green function for the \emph{Dirac fermions}  
(note the factor of 2 because of the two 
fermion generations) is 
\beq\label{eq:FProduction}
\im\ \tilde \Gamma^{(2)}(\chi) = \frac{y^2}{4\pi} \chi \left( 1 - \frac{4 m_f^2}{\chi}\right) \Theta (\chi - 4 m_f^2)~. 
\eeq
A small subtlety is that due to the mixings resulting from the couplings~\eqref{eq:DarkFermionL} we are producing 
Majorana rather than Dirac fermions. However, if we are interested in total production of all the available states, the final 
result will still be given by expression~\eqref{eq:FProduction}. 

After picking up together all the relevant terms, one finds the produced number of particles per comoving  volume 
\beq
Y_\alpha = \frac{225}{4 \pi^2 } \frac{1}{\sqrt{g_*} T M_{Pl}} \left( \frac{\beta}{H}\right) \left(\frac{\cE}{A} \right)~,
\eeq
with $\beta^{-1}$ being the duration of PT, typically $\beta \sim (100 \ldots 1000)\times  H$~\cite{Hogan:1984hx}.

We show the expected abundances for various points in the parameter space on Fig.~\ref{fig:abundances}. 
The maximal possible $\gamma$ that we can expect is 
\beq
\gamma \lesssim \left( \frac{\beta}{H}\right)^{-1} \frac{M_{pl}}{f}~, 
\eeq
namely can potentially be as large as $(10^{14} \ldots 10^{15})$. In practice this value is smaller than the maximum 
possible velocity, because this naive estimation neglects the $\log \gamma$ friction terms, which might also become
important for these values of $\gamma $. However, even for relatively mild values of $\gamma\sim 10^9$ we can get 
a correct baryonic abundance with pretty low CP-efficiency, in fact as small as $\epsilon_{CP} \sim 10^{-5}$
which can very easily achieved perturbatively without any need invoking resonant baryogenesis.


\section{Experimental Signatures}
\label{sec:signatures}
Because the entire mechanism is crucially based on the strong first order PT at relatively low temperature with runaway 
bubbles, the prediction of the primordial gravity waves is completely model independent. Primordial gravitational 
waves are generic signatures of the strong cosmological first order PT, and their effects were studied in detail
in the context of the non-standard 
EWPT~\cite{Chala:2016ykx,Grojean:2006bp,Huber:2007vva,Delaunay:2007wb,Espinosa:2008kw,Das:2009ue,Addazi:2016fbj}, 
in the case of SM QCDPT with the lepton asymmetry~\cite{Caprini:2010xv}, and, recently, in the case of low-temperature hidden 
sector PT~\cite{Schwaller:2015tja,Jaeckel:2016jlh}. The latest case is in fact similar to our study. 

We analyze the predicted Gravity
waves in the first subsection of this section. We later turn to comment on more model dependent signature, like 
$n - \bar n$ oscillations, double nucleon decay and even possible collider signatures. 

\subsection{Gravitational Waves}
The gravitational wave signal is especially interesting and intriguing in this particular mechanism not only 
because that it is very generic, but also because of its potential detectability in the forseeable future. Indeed, 
strong cosmological phase transitions at temperatures in range $(1 \ldots 50)$~GeV can be precisely in the 
preferred range of the future eLISA detector. At this point it is little bit 
difficult to estimate the sensitivity of the eLISA exactly, because the sensitivity studies are now under way. However
lots of future designs of the eLISA 
will be  able either provide evidence to our mechanism, or falsify
big parts of its parameter space.  

During a strong first order phase transition the primordial gravitational wave is produced due to three different processes 
which all generically coexist: collisions of bubble walls, sounds wave in the plasma after the bubbles collided and
magnetohydrodinamic (MHD) turbulence in plasma forming after the bubbles collided. Therefore 
\beq
\Omega_{GW} h^2 = \Omega_\varphi h^2 + \Omega_{sw} h^2 + \Omega_{turb} h^2~,  
\eeq   
where $\Omega_\varphi$ stands for the first mentioned contribution. In our following analysis we will closely 
follow the discussion of~\cite{Caprini:2015zlo}, from where we also take the expressions for all these distributions. We provide 
here only the final formula for our estimations, and for more details the interested reader is referred to~\cite{Caprini:2015zlo}
and references therein. 

The fundamental parameter, which largely determines the intensity of the gravitational wave emitted by 
the strong 1st order phase transition is the ratio between the vacuum energy density and the radiation energy 
density during the PT:
\beq
\alpha \equiv \frac{\rho_{vac}}{\rho_{rad}} = \frac{30}{\pi^2} 
\frac{V_{th}(0, T_{nuc}) - V_{th}(f(T_{nuc}); T_{nuc}) }{g_*  T^4}
\eeq  
With this parameter and the duration of the PT $\beta^{-1}$ we can express the intensity of the gravity waves as 
a function of the frequency.
Note, 
that the discussion in Ref~\cite{Caprini:2015zlo} is pretty generic and applies to different velocities of the bubble walls. Here 
we assume runaway bubble walls, $v_w = 1$, in which case the relevant expressions for intensity are:
\beq
\Omega_\varphi h^2 (f) & = & 0.13 \times 10^{-5} \left( \frac{H}{\beta}\right)^2 
\left(\frac{\kappa_\varphi \alpha}{1 + \alpha} \right)^2 \left( \frac{100}{g_*}\right)^{1/3} S_{env(f)}~,\\
\Omega_{sw} h^2 (f) & = & 2.65 \times 10^{-6} \left( \frac{H}{\beta} \right) \left(  \frac{\kappa_v \alpha}{1+ \alpha}
\right)^2 \left( \frac{100}{g_*}\right)^{1/3} S_{sw} (f)~, \\
\Omega_{turb} h^2 (f)  & = & 3.35 \times 10^{-4} \left( \frac{H}{\beta} \right) \left(  
\frac{\kappa_{turb} \alpha}{1+ \alpha}
\right)^{3/2} \left( \frac{100}{g_*}\right)^{1/3} S_{turb}(f)~.
\eeq
The functions $S(f)$, which determine the dependence of the gravitational wave intensity on the frequency are:
\beq
S_{env}(f) & = & \frac{3.8 (f / f_{env})^{2.8}}{1 + 2.8 (f / f_{env})^{3.8}}~, \\
S_{sw}(f) & = & \left( \frac{f}{f_{sw}}  \right)^3 \left( 
\frac{7}{4 + 3 (f / f_{sw})^2}\right)^{7/2}~,\\
S_{turb}(f)  &= &\frac{(f/f_{turb})^3}{\left( 1 + (f / f_{turb})\right)^{11/3} 
(1 + 8\pi f/h_*)} \eeq
with
\beq\label{eq:redshift}
h_* = \left( \frac{T}{100~\gev}\right) \left( \frac{g_*}{100} \right)^{1/5}\, 
1.65 \times 10^{-5}~{\rm Hz}~. 
\eeq
In these expressions the frequencies $f_{env}$, $f_{sw}$ and $f_{turb}$ are the frequencies at which each 
of the functions peaked. In general we expect them to be of order $\sim \beta $ (inverse duration of the phase transition) 
and further redshifted with $h_*$, given by Eq.~\eqref{eq:redshift}. Taking all this into account we get the expected 
peak frequencies:
\beq
f_{env } & = & \left(  \frac{\beta}{H}\right) \left( \frac{T}{100~\gev} \right) \left( \frac{g_*}{100} \right)^{1/6} \,
3.8 \times 10^{-6}~{\rm Hz}\\
f_{sw} & = & \left( \frac{\beta}{H}\right) \left( \frac{T}{100~\gev}\right) \left( \frac{g_*}{100}\right)^{1/6}\,
1.9 \times 10^{-5}~{\rm Hz}\\
f_{turb} & = & \left( \frac{\beta}{H}\right) \left( \frac{T}{100~\gev}\right) \left( \frac{g_*}{100}\right)^{1/6}\,
2.7 \times 10^{-5}~{\rm Hz}
\eeq 

\begin{figure}
\centering
\includegraphics[width = .49\textwidth]{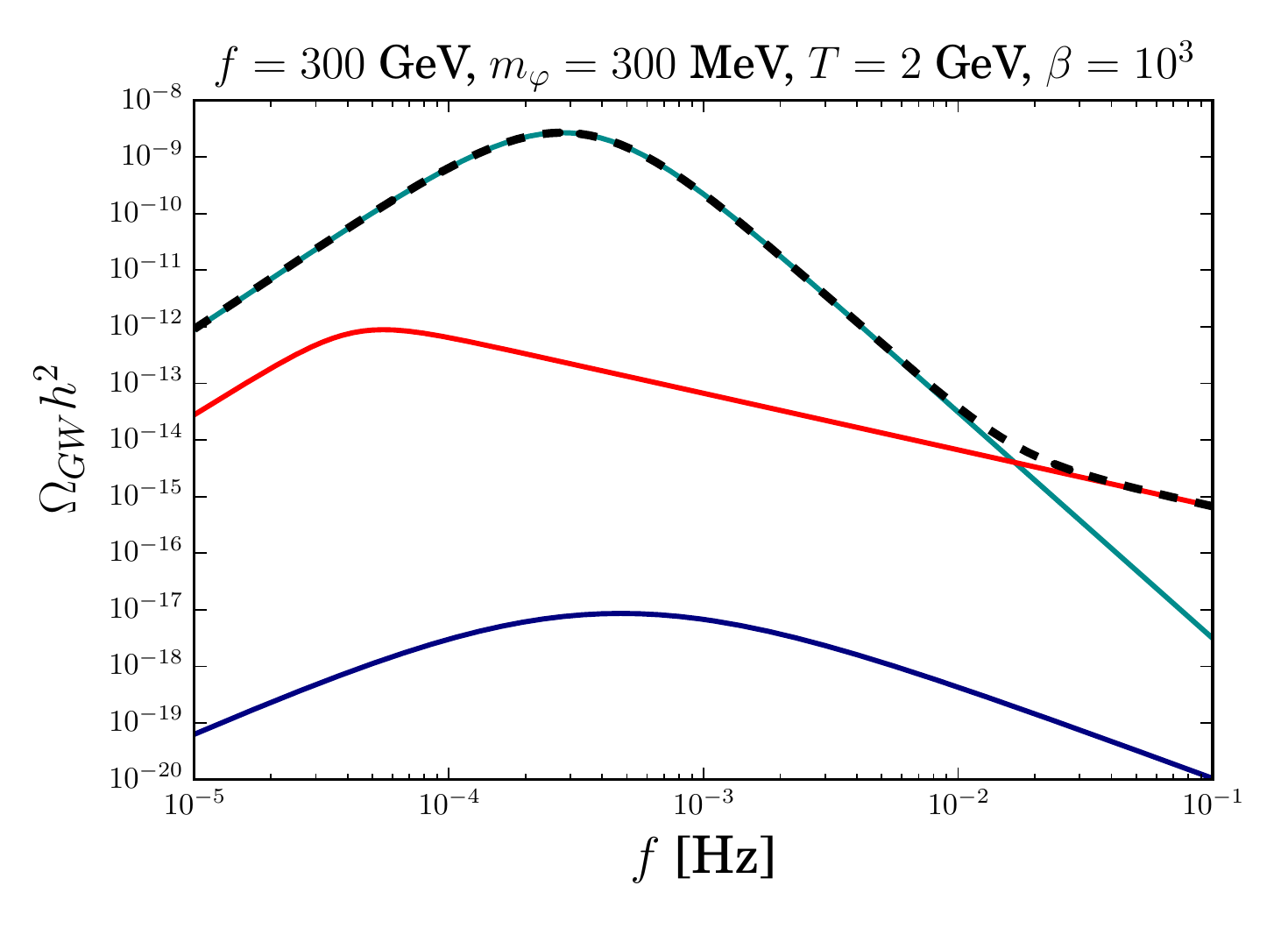}
\includegraphics[width = .49\textwidth]{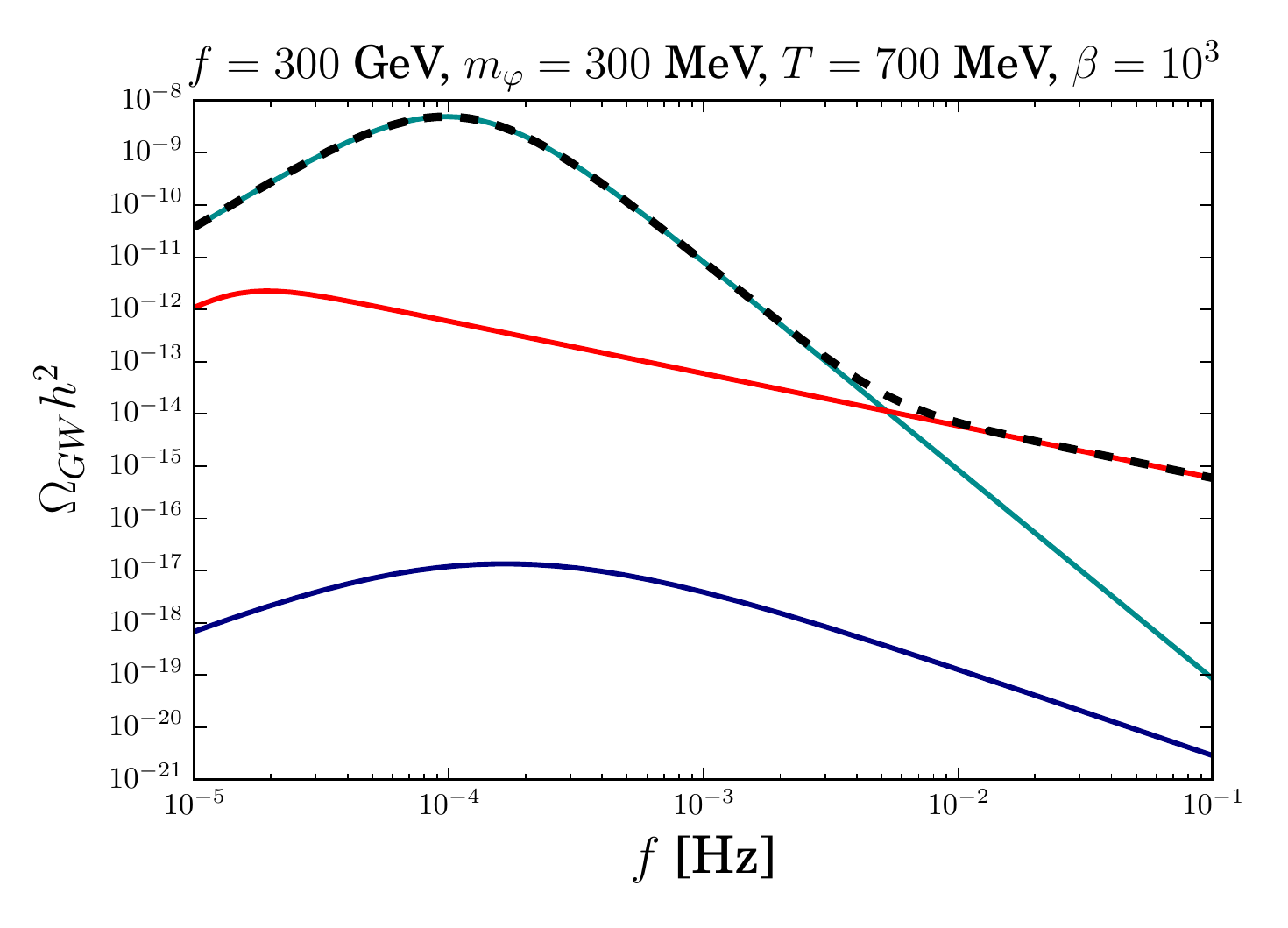}
\includegraphics[width = .49\textwidth]{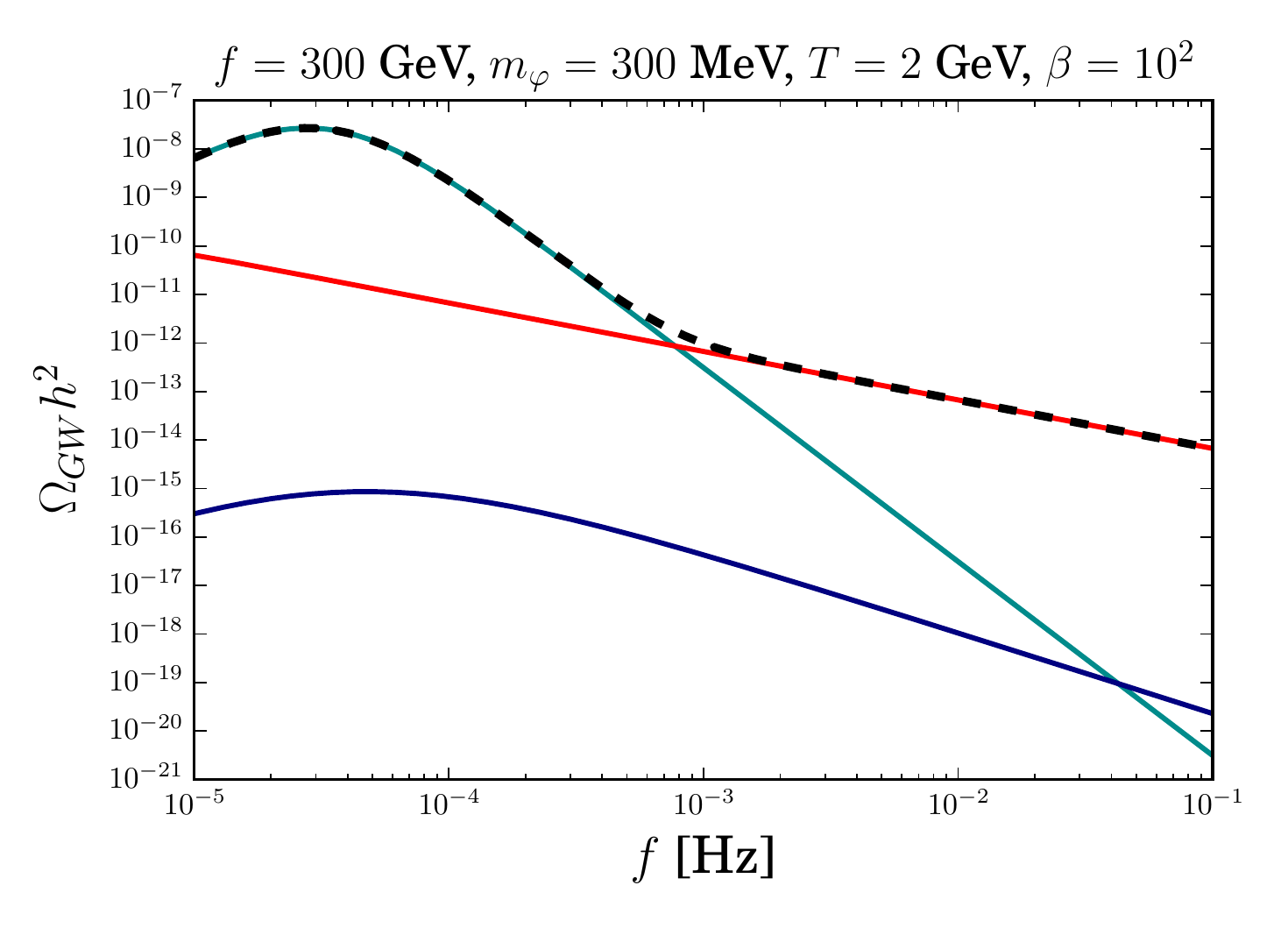}
\includegraphics[width = .49\textwidth]{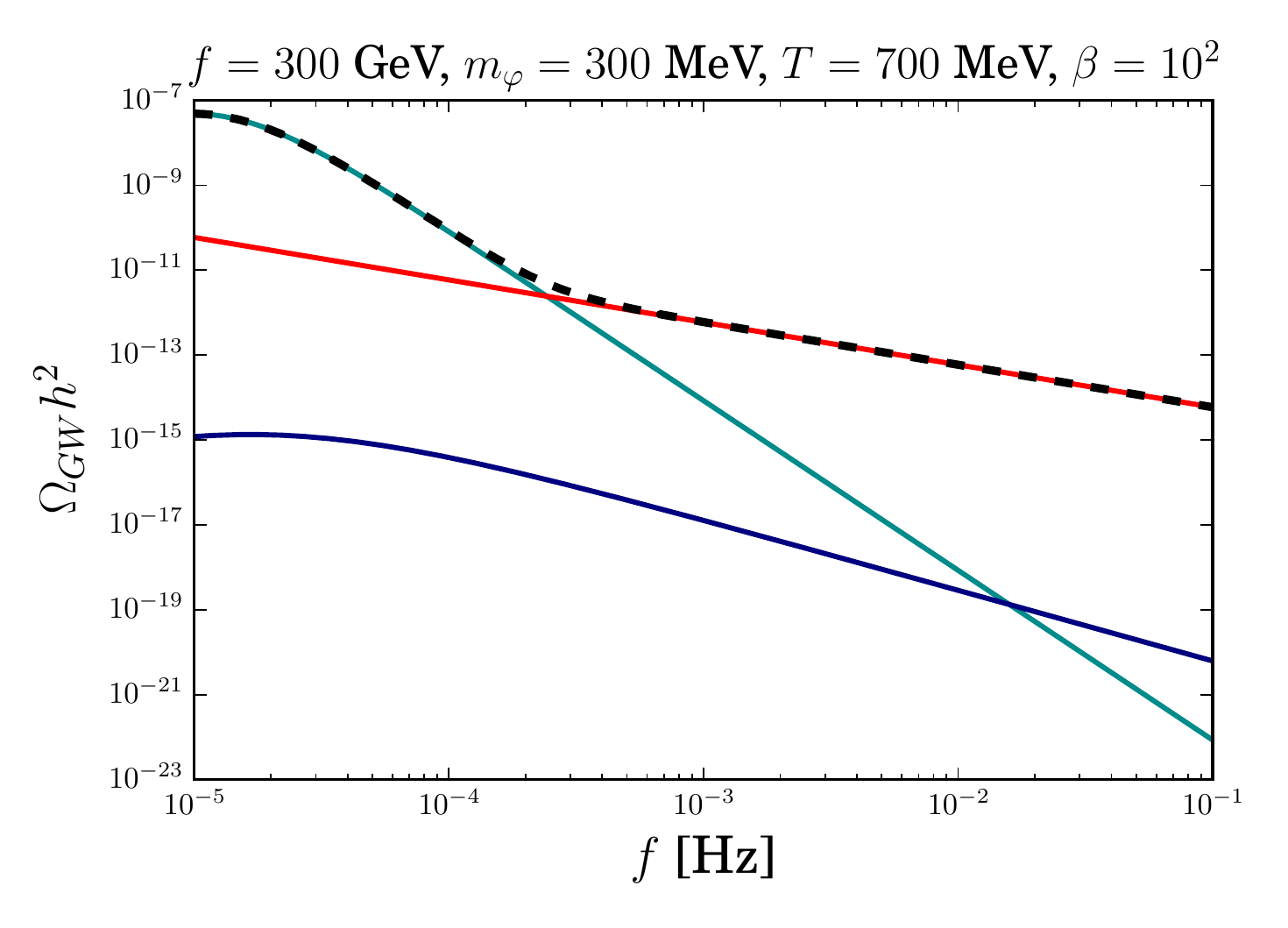}
\caption{Gravitational wave spectrum of one particular point point in parameter space of a model, described
in Sec.~\ref{sec:su2model}. On the left panels we assume $T_{nuc} = 2$~GeV, just below the bound on the nucleation
temperature that we derive in Subsec.~\ref{subsec:runaway}. On the right panel we assume $T_{nuc} = 0.7$~GeV,  
significantly below the upper bound. On the upper panels we show the spectrum for relatively high $\beta$, and 
lower its value on the lower panel. The dark blue line stands for the MHD turbulence contribution, the red line stands 
fro the direct contribution from bubble collisions $\Omega_\varphi h^2$ and the cyan line stands for the sound waves 
contribution. The total intensity is indicated by the dashed black line. Everywhere $\epsilon = 0.05$ is assumed.}
\label{fig:GW}
\end{figure}

Finally the parameters $\kappa_{env,\, v,\, turb} $ stand for the fraction of the vacuum energy  converted into the gradient 
energy of the dark higgs field, bulk motion of the fluid and into the MHD turbulence respectively. These efficiencies for the bulk 
motion and turbulence in $v_w \sim 1$ limit are given by 
\beq
\kappa_v & = & \alpha (0.73 + 0.083\sqrt{\alpha} + \alpha)^{-1}, \ \ \ \kappa_{turb} = \epsilon \kappa_v~, 
\eeq
where $\epsilon$ is of order $0.05 \ldots 0.1$ based on the numerical simulations. The efficiency for the gradient 
energy of the higgs is 
\beq
\kappa_{env} = \frac{\alpha - \alpha_{\infty}}{\alpha} \ \ \ {\rm where} \ \ \ \alpha_\infty =
\frac{30}{24 \pi^2} \frac{\sum_a c_a (m_a^2(f(T)) - m_a^2(0)}{g_*(T) T}~,
\eeq
with $c_a = N$ (number of degrees of freedom) for the bosons and $c_a = N/2$ for the fermions. 
Note also that condition of~\cite{Espinosa:2010hh} for the runaway bubbles that we have 
already mentioned in Sec.~\ref{sec:mechanism}
is exactly $\alpha > \alpha_\infty$, namely that non-negligible amount of energy density is damped into the gradient 
energy of the higgs field.  

Because full analysis of the parameter space of the toy model is well beyond the scope of our work, we do not 
study here, which parts of the parameter space can be explicitly probed by eLISA (given, also, the uncertainties 
in future eLISA sensitivity). But to illustrate the point we plot on Fig.~\ref{fig:GW} 
the signal expected spectrum of one of the points in the parameter space, that we have already analyzed explicitly 
in Subsec.~\ref{subsec:runaway}. 

Note that all these points that we illustrate on Fig.~\ref{fig:GW} can be well within the reach of eLISA, at least 
if we rely on the old LISA configuration.\footnote{This corresponds to 6 links, 5 million~km arm length and duration of 
5~years. For more details on possible configurations of eLISA and the expected noise the reader is 
referred to~\cite{Caprini:2015zlo} and especially to~\cite{Klein:2015hvg}.} 
Not surprisingly, the intensity of the signal is peaked at very low frequency
and the highest intensity point might  be inaccessible  to eLISA. 
This is especially true for low nucleation temperature and low $\beta$, such that  
the intensity distributions are peaked at frequencies as low as $10~\mu$Hz, where eLISA almost looses its sensitivity. 
However, 
due to very big energy release in the PT  in this region, the signal from the tail is strong enough to yield 
$\Omega_{GW} h^2 \sim (10^{-13} \ldots 10^{-12})$ for frequencies of order 10~mHz, where the eLISA sensitivity is 
maximized. 
The $\alpha$
gets values which are as high as 500 for $T \approx 0.7$~GeV.  The most optimistic eLISA sensitivity projected sensitivity
will be able to probe in this region $\Omega_{GW} h^2$ as low as $10^{-14}$, while its sensitivity drops to values 
as low as $\Omega_{GW} h^2 \sim 10^{-10}$ for frequencies of order 10~$\mu$Hz.  
The contribution to the GW intensity 
is heavily dominated by the sounds waves in the low frequency regions and by the direct bubble collisions contribution 
at higher frequencies, while the MHD turbulence is always negligibly small.   

\subsection{Brief comments on other possible signatures}
It is tempting to conclude, based on the interactions~\eqref{eq:FermionInteractions} that the model that we have introduced, 
and largely the entire mechanism, inevitably predict $n - \bar n$ oscillations and the respective bounds apply. Indeed, by taking 
the first term of the equation~\eqref{eq:FermionInteractions} on its face value and integrating out the heavy sterile fermion 
$e_\alpha $ one gets 
\beq
\cL \supset \frac{\lambda _{\alpha i j k} \lambda^\alpha_{lmn} u^c_i d^c_j d^c_k u^c_l d^c_m d^c_n}{m_e \Lambda^4}
\eeq
which looks like a perfect $n - \bar n$ oscillations operator. Nonetheless, the color contraction in this expression 
forces antisymmetrization on the flavor indices in the down-sector. Namely, at least two of the down-type quarks must be either 
$s$ or $b$-quarks. We get a qualitatively different behavior if we consider the second operator in the~\eqref{eq:FermionInteractions}, 
namely the $QQ(d^c)^\dagger$. Here we must antisymmetrize both on the color and the $SU(2)$ indices of the $Q$ fields, and 
therefore there is no problem with getting a non-zero coefficient for the same generations fields. 

Although the $n - \bar n$ operator does not form at the leading order, if only the coupling $\lambda$ is invoked, 
with democratic flavor structure of the coupling and 
sizable $\lambda_{\alpha 112}$ one does get a double nucleon  decay $pp \to KK$. The lifetime of this process is 
constrained to be bigger than $1.7 \times 10^{32}$~years~\cite{Litos:2014fxa}, which was inferred from non-observation 
of  $^{16}O$ nucleus decay  into the  $^{14}C K^+ K^+$. Rephrasing the calculation of~\cite{Goity:1994dq} in our 
variables we get 
\beq
\tau \approx \frac{\pi}{8} \frac{\Lambda^8 m_e^2 m_N^2}{ \rho \Lambda_{QCD}^{10}}~,
\eeq
where $\rho \approx 1.91\times 10^{-3}~\gev$ is a nuclear matter density and $\Lambda$ is the suppression scale 
from Eq.~\eqref{eq:FermionInteractions} assuming that $\lambda = 1$. In fact, one gets precisely the same expression 
for the lifetime if $\eta = 1$ is assumed. This translates to the bound in the range of PeV scale, and in fact 
it is the most stringent bound that one can obtain even in the case of the unsuppressed $n - \bar n$ oscillations rate. However, as we 
discussed in Sec~\ref{sec:su2model}, it is not difficult to satisfy this bound by taking the coupling of the 
exotic colored scalar to the $d^c$
to be of order $\cO(10^{-3})$ or smaller for $m_\Delta \sim $~TeV. 
In lots of senses prediction of the dinucleon decay is an interesting feature of a model, 
that we presented, but the exact scale unfortunately depends  on the model details and cannot be predicted firmly, also because the 
exact flavor structure of the new operators are not known, and the couplings to the light generations can in principle 
be highly suppressed. 

Let us now again consider the $u^c d^c d^c$ coupling. 
The contribution to the $n - \bar n $ oscillations is of course not identically zero, and, if the coupling to the strange quark 
is non-zero, one gets it from the strange quark component of the neutron, which is of order 1\%. In order to estimate this constraint
we rely on the estimations of~\cite{Babu:2006wz}. Although the model of baryogenesis that that reference proposes is very 
different, the fermion sector shares some clear similarities with ours and the estimation of $n - \bar n$ oscillation rate 
goes along the same lines. With the bound on the neutron 
oscillation time $\tau_n > 1.3 \times 10^8$~years~\cite{Chung:2002fx}
and assuming that  $\lambda_{\alpha 112} \sim \cO(1)$ we get
\beq
m_e \Lambda^4 \gtrsim 10^{19}~\gev^5~.
\eeq
For $m_e \sim 1$~TeV, this translated to a very minor bound $\Lambda \gtrsim 10$~TeV, very subdominant to the double nucleon 
decay. 

In case of the $QQ(d^c)^\dagger$ coupling, assuming that  one have $\eta_{111} \sim \cO(1)$, one gets a bound from 
the $n-\bar n$ oscillations in the vicinity of 100~TeV. Note though that in this case it does not directly compare to the dinucleon decay, 
because different flavor operators are responsible for these processes. 

Finally, let us make a brief comment about the collider phenomenology. Although the mechanism makes absolutely no firm 
prediction in this area, one can still expect the the bosonic operators in the hidden valley couple to the SM higgs operator 
$|H|^2$, because it is the lowest dimension SM gauge invariant scalar operator.  This can lead to the rare exotic higgs decays, 
and intriguing possibility, which has been studied both in theoretical and experimental context~\cite{Curtin:2013fra}. In the model
that we presented here such an operator is Eq~\eqref{eq:HiggsPortal}. Although in our case the invisible rate is way too small 
to be observed at the LHC, it might still be visible in the future leptonic and hadronic colliders. And, not unlikely, in other 
models which take advantage of our mechanism, the rate can be even bigger. 

Another potentially interesting aspect of the collider phenomenology has to do with the production and decay of the colored
scalars $\Delta$. Of course the existence of these particles is strongly model-dependent, however, if exist, they can be abundantly 
produced by the LHC. Even though these particles are colored, the LHC potential to discover them is modest, mainly because 
they decay into pair of anonymous jets, similar to the RPV stops in SUSY with the baryon number violation. 
The constraints on these particles are exactly the same as the constraints on the BNV stops. The ATLAS 
search~\cite{Aad:2016kww} bounds the mass of this particle to be above $\gtrsim 320$~GeV in the case that one 
of the resulting jets is b-tagged, and puts no constraints in case the decay proceeds into the light jets. The companion 
CMS search~\cite{Khachatryan:2014lpa} is also looking for the resonances in a decay channel, where one of the final state
jets is heavy, and they constrain such particles in the mass window $200~\gev < m_{\Delta} < 380~\gev$. In the most general case, 
if $\Delta$ decays into a pair of light jets only CDF bound applies~\cite{Aaltonen:2013hya} $m_\Delta > 100$~GeV. 
In general, it is a notoriously hard search, mostly due to the trigger issues and even after the 13~TeV run the bound is not
expected to exceed 600~GeV~\cite{Bai:2013xla}.

\section{Conclusions and Outlook}
\label{sec:conc}
In this paper we have introduced a novel mechanism to produce the Universe baryonic asymmetry via out-of-equilibrium decays 
of heavy particles which are produced non-thermally. The mechanism can work at the temperatures that are much lower
than the masses of the decaying particles and therefore there is conceptually no need for the Universe to be reheated 
after inflation up to the masses of the decaying particles.   We have also presented a simple model that serves an existence 
proof that the mechanism can be embedded into a realistic framework, which satisfies all the known cosmological, 
collider and other constraints.  

Although we have concentrated on the low energy regime and post-sphaleron baryogenesis, even this part of parameter 
space does not necessarily provide us with robust LHC signatures. In fact, one might hope for such signatures 
(like long-living color triplet at the TeV scale), however they all appear to be highly model-dependent. On the other hand, 
the prediction of primordial gravitational waves background is model independent, because it relies  on the crucial component 
of the mechanism: low-energy strong first order cosmological PT with the runaway bubble walls. This opens intriguing 
perspectives for the future eLISA interferometer, which can potentially provide first supportive evidence of this mechanism. 

There are still several open questions on the theory side which should still be answered to address the feasibility 
of our mechanism. We have mentioned them already in the text, but it would probably be useful to summarize
them here.  Our mechanism relies on the possibility of efficiently producing heavy states during bubble collisions. This requires
large value of the Lorentz factor $\gamma$. We have made use of the fact that such large values are potentially achievable
during a PT in which bubbles runaway. The exact criterion under which
runaway bubbles are present is still under debate we have  decided to
stay on the safe side by adopting the most conservative criterion. We
have not computed ourselves the value of $\gamma$ achieved during the
PT, but rather indicated an upper bound for which our mechanism can
work. Of course, a crucial step would be to quantitatively check if
such large values of the Lorentz boost can be achieved. Also we note, that we have not analyzed this theory in the limit 
of resonant baryon asymmetry production. This might be interesting to do in order to understand whether lower 
values of $\gamma $ can be relevant for our mechanism. 
Finally, a
more thorough, most probably numerical, study will be required to
estimate the role of elasticity during bubbles collisions and the
 efficiency of particle production.

Our proposal also opens a handful of new directions in the studies of the Universe baryon asymmetry and leaves several 
interesting questions open? Are there any \emph{natural} scenarios that can accommodate the mechanism that we 
propose? For example, can it be confinement PT, which does not require a fine-tuned higgs mechanism? Or whether 
this mechanism can emerge in a hidden sector theory which is naturally higgsed, either by virtue of the hidden 
valley being naturally supersymmetric or because the higgs is emergent (composite higgs scenario)?   Moreover, 
in this paper we have just brought one model, kind of existence proof that the mechanism can work, but it is 
still unclear how generic the mechanism is. It would be interesting to see more models along these lines, 
understand their parameter space, and more important see what are the experimental implications of these models.

\acknowledgments{We are very grateful to T.~Konstandin, G.~Nardini, J.~M.~No  
for useful discussions. We are also very grateful to T.~Konstandin for his helpful 
comments on the manuscript. We are also grateful to the anonymous referee for his numerous 
useful comments and for pointing us out that the lifetime of the dark gauge bosons is much shorter 
than we have estimated in the previous version. The research of A.R is 
supported by the Swiss National Science Foundation (SNSF), project {\sl Investigating the
Nature of Dark Matter}, project number: 200020-159223.  
The research of AK was partially supported by the Munich Institute for Astro- and Particle Physics (MIAPP) 
of the DFG cluster of excellence "Origin and Structure of the Universe}  

\bibliography{refs}
\bibliographystyle{JHEP}
\end{document}